%% file: Main.tex
\documentclass[twocolumn]{NobArticle}
\usepackage{subcaption} 
\usepackage{svg}
\usepackage{amsmath} 
\usepackage{graphicx}
\usepackage{booktabs} 
\usepackage{caption}
\usepackage{flafter}
\usepackage{graphicx}
\usepackage{hyperref}
\usepackage{xcolor}
\usepackage{multirow}

\title{Uni-AIMS: AI-Powered Microscopy Image Analysis}

\author{
    Yanhui Hong$^1$\thanks{\scriptsize Equal contribution. For a full description of each author’s contribution, see Section Author Contributions.} \thanks{\scriptsize Corresponding author.}~, 
    Nan Wang$^1$\footnotemark[1],
    Zhiyi Xia$^1$\footnotemark[1],
    Haoyi Tao$^1$,
    Xi Fang$^1$,
    Yiming Li$^1$,
    Jiankun Wang$^1$,
    Peng Jin$^1$,
    Xiaochen Cai$^1$,
    Shengyu Li$^1$,
    Ziqi Chen$^1$,
    Zezhong Zhang$^2$,
    Guolin Ke$^1$,
    Linfeng Zhang$^{1,2}$
}

\date{
    \quad $^1$DP Technology, Beijing, 100080, China. \\ 
    \quad $^2$AI for Science Institute, Beijing 100080, China. \\ 
}    

\renewcommand{\maketitlehookd}{%
\begin{abstract}
     \noindent This paper presents a systematic solution for the intelligent recognition and automatic analysis of microscopy images. We developed a data engine that generates high-quality annotated datasets through a combination of the collection of diverse microscopy images from experiments, synthetic data generation and a human-in-the-loop annotation process. To address the unique challenges of microscopy images, we propose a segmentation model capable of robustly detecting both small and large objects. The model effectively identifies and separates thousands of closely situated targets, even in cluttered visual environments. Furthermore, our solution supports the precise automatic recognition of image scale bars, an essential feature in quantitative microscopic analysis. Building upon these components, we have constructed a comprehensive intelligent analysis platform and validated its effectiveness and practicality in real-world applications. This study not only advances automatic recognition in microscopy imaging but also ensures scalability and generalizability across multiple application domains, offering a powerful tool for automated microscopic analysis in interdisciplinary research. A \href{https://www.bohrium.com/apps/uni-aims}{\textcolor{cyan}{online application}} is made available for researchers to access and evaluate the proposed automated analysis service.

    \small{\textbf{Index Terms:} Microscopy Image Analysis, Deep Learning, Instance Segmentation, Microstructure Characterization}
\end{abstract}
}

\begin{document}

\small
\maketitle
\input{Sections/01-Introduction}
\input{Sections/02-Architecture}
\input{Sections/03-Data_Engine}
\input{Sections/04-Model}
\input{Sections/05-Platform}
\input{Sections/06-Application}

\input{Sections/07-Conclusion}

\section*{Author Contributions} \label{sec:authors}

Yanhui Hong led the project, constructed the dataset, explored instance segmentation models on microscopy image datasets, and wrote the manuscript.
Nan Wang implemented data generation, conducted research on dense particle segmentation, and contributed to the manuscript writing.
Zhiyi Xia implemented and accelerated the instance segmentation model and participated in the manuscript writing.
Haoyi Tao and Xi Fang implemented automatic recognition of image scale bars.
Yiming Li conducted research on data generation.
Peng Jin and Jiankun Wang built the intelligent analysis platform.
Xiaochen Cai, Shengyu Li, and Ziqi Chen jointly developed the annotation platform and are in charge of the management and implementation of manual annotation processes.
Zezhong Zhang, Guolin Ke and Linfeng Zhang provided technical guidance and contributed to the manuscript revision.

\printbibliography

\end{document}

%% file: Sections/01-Introduction.tex
\section{Introduction}

\subsection{Background}

In various fields of scientific research and industrial applications, precise analysis of microstructures has become a crucial factor driving technological advancements and expanding the boundaries of knowledge (\cite{murphy2012fundamentals}). From the development and optimization of new materials to the in-depth exploration of disease pathogenesis, from the detailed observation of cellular biology to the performance evaluation of nanoscale devices, microscopy images contain a wealth of valuable information. Efficient and accurate interpretation of these images is of irreplaceable strategic significance for accelerating scientific progress, improving production efficiency, and promoting interdisciplinary integration.

In industrial applications, Scanning Electron Microscopy (SEM) is widely used to characterize material morphology, including particle size, shape, distribution, and surface roughness. These morphological features are critical for the standardization, quality assurance, and performance optimization of materials and products (\cite{goldstein2017sem}). In scientific research, microscopy images reveal microstructural details such as nanoparticles, crystal defects, and pore structures, linking these features to macroscopic properties like mechanical strength and electrical conductivity. Such insights facilitate the understanding of fundamental mechanisms and the development of advanced materials (\cite{callister2018materials}).

\subsection{Limitations of Traditional Methods}

However, traditional microscopy image analysis methods (\cite{collins2007imagej}) face numerous challenges. As research complexity increases and application scenarios expand, the vast, diverse, and continuously growing microscopy datasets make manual analysis approaches increasingly inadequate. Manual analysis is not only inefficient but also highly subjective, making it difficult to meet the demands for rapid and objective decision-making.

Moreover, different disciplines are imposing increasingly specific requirements on microscopy image analysis. For example, in energy materials research, precise analysis of pore density and particle size distribution is critical for performance evaluation; in metal processing and additive manufacturing, grain size grading and defect detection directly affect product quality and reliability. Traditional analysis methods often lack consistency and generalizability when addressing these complex tasks, making efficient knowledge transfer and cross-domain application difficult.

\subsection{Opportunities with Machine Learning}

In recent years, with the rapid development of machine learning and deep learning techniques in the fields of image recognition and analysis (\cite{redmon2016lookonceunifiedrealtime}, \cite{wu2019detectron2}, \cite{kirillov2023segany}), microscopy image analysis has encountered new opportunities (\cite{choudhary2022recent}). Data-driven approaches can automatically learn image features, significantly improving the efficiency and accuracy of recognition and analysis. Existing studies have attempted to apply cutting-edge deep learning models to microscopy image analysis, achieving promising results, particularly in medical imaging (\cite{MedSAM}, \cite{archit2025segment}) and cellular biology (\cite{moen2019cellimage}).

Several studies have explored the application of machine learning models for analyzing microscopy images in materials science. For example, Yildirim et al. (\cite{yildirim2021bayesian}) proposed a Bayesian deep learning model for semantic segmentation and localization of particle instances in electron microscopy (EM) images, while also estimating uncertainty. In another study, Shi et al. (\cite{shi2022automatic}) employed U-Net and MultiResUNet architectures for semantic segmentation of SEM images of microplastics, combined with a fine-tuned VGG16 network for shape classification. Zhang et al. (\cite{zhang2019spherical}) applied a Mask R-CNN-based instance segmentation framework to identify spherical nanoparticles in SEM images. Similarly, Bals et al. (\cite{bals2023deep}) used a UNet++ architecture and introduced a pixel-wise loss map based on distance and intensity, encouraging the network to better distinguish between densely packed particles.

While these studies demonstrate the potential of deep learning for analyzing microscopy images in materials science, they are typically restricted to specific imaging modalities or narrowly defined sample types. Consequently, they lack the generalizability necessary to address the wide variety of styles, content, and object distributions present across broader categories of microscopy images. This limitation stems largely from the unique characteristics of material microscopy data: diverse imaging techniques and specimen types lead to substantial variations in resolution, contrast, texture patterns, and background noise. Moreover, the number of object instances within a single image can vary drastically—from just a few to tens of thousands—accompanied by significant differences in object size and frequent occurrences of particle stacking and clustering. These complexities pose considerable challenges for directly applying models developed for natural or biomedical imagery to the analysis of materials microscopy images.

Furthermore, compared to the extensive datasets available in medical (\cite{kus2024medsegbench}) and biological fields (\cite{goodman2018datasciencebowl}), publicly accessible high-quality microscopy image datasets in materials science are extremely scarce, failing to meet the data volume requirements for training deep learning models. Additionally, annotation of materials microscopy images demands specialized domain knowledge and is both time-consuming and labor-intensive; the lack of high-quality annotated data further restricts the development and application of intelligent analysis methods.

\subsection{Our Work}

To address the aforementioned challenges, this study proposes a systematic solution for the intelligent recognition of microscopy images. The framework comprehensively covers the entire pipeline, from data acquisition and the construction of a multi-source annotation system to deep model training and multi-task applications, ultimately integrating into a powerful and user-friendly intelligent analysis platform.

Considering the unique characteristics of microscopy images, we propose an instance segmentation model that fully accounts for texture features and morphological diversity. By leveraging advanced deep learning algorithms and architectural innovations, the model is capable of accurately detecting targets across scales and robustly identifying and separating thousands of densely packed instances within cluttered backgrounds.

This research effectively overcomes key challenges in materials microscopy image analysis, including data scarcity, annotation difficulties, and sample diversity, achieving both high efficiency and strong generalizability across different application domains. The proposed intelligent platform provides a robust automated tool for microstructural analysis across multiple disciplines, aiming to meet the urgent needs of researchers and industrial practitioners for efficient and precise microstructural analysis.

In the following chapters, we will provide a detailed description of each component of the framework and its performance in practical applications.

%% file: Sections/02-Architecture.tex
\section{Framework Overview}

\begin{figure*}[htbp]
    \centering
    \includegraphics[width=0.9\textwidth]{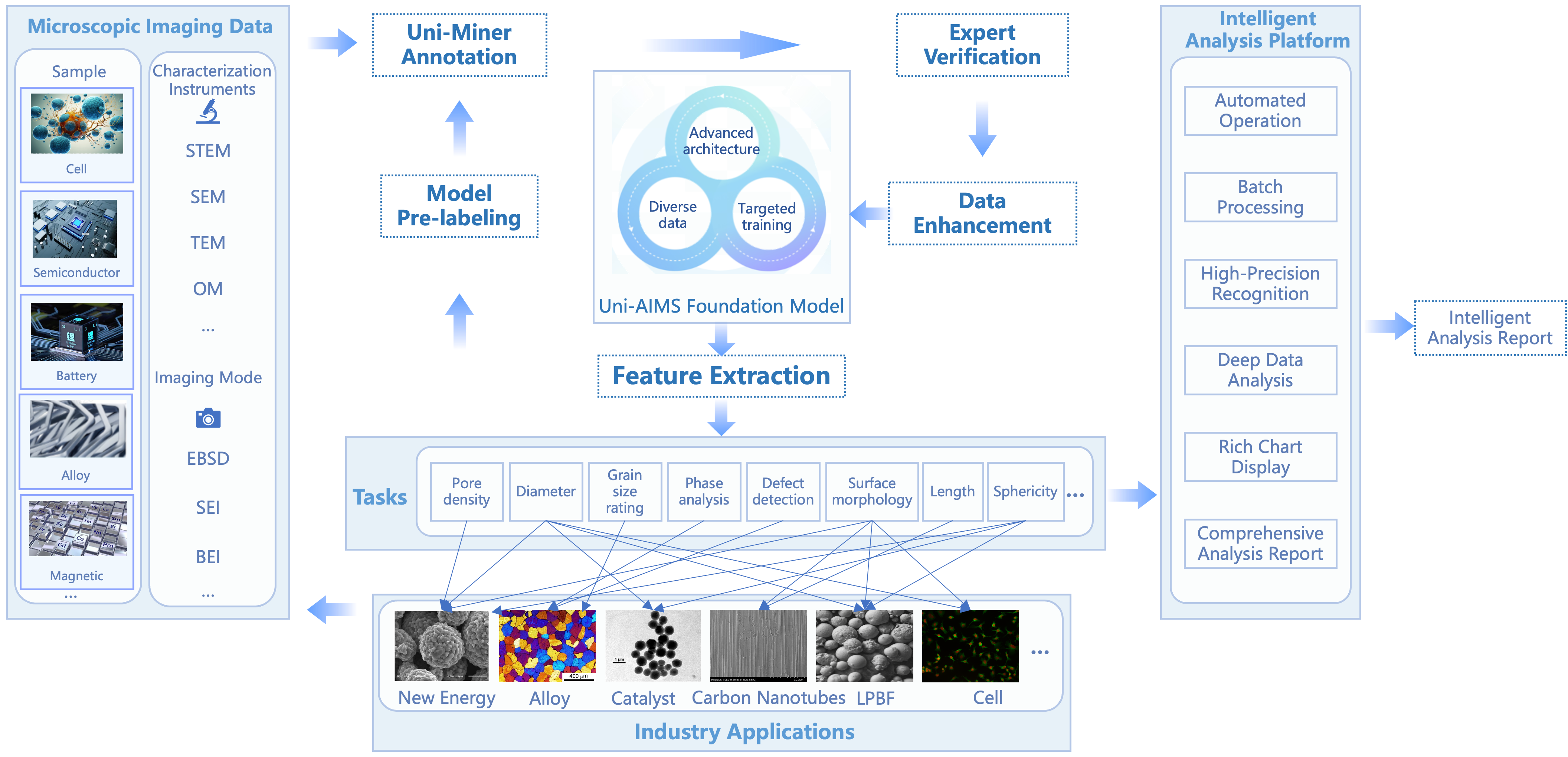}
    \caption{The proposed microscopy image analysis framework. On the left, microscopy images are acquired from various sources and labeled via a human-in-the-loop approach. In the middle, the labeled data undergoes data augmentation to train the Uni-AIMS foundation model for feature extraction. Below, different analysis tasks and their industry applications are showcased. On the right, our intelligent analysis platform automatically generates reports. Throughout this process, data acquisition, labeling, and model training form a data flywheel that iteratively improves the system.}
    \label{fig:framework}
\end{figure*}

We have developed a systematic research framework for intelligent recognition and analysis of microscopy images. This framework encompasses data acquisition, construction of a multi-source annotation system, deep learning model training, and multi-task applications, ultimately integrated into a unified intelligent analysis platform. The goal is to address the pressing need for automated microstructural analysis across various scientific and engineering disciplines. Figure~\ref{fig:framework} illustrates the core components of this framework and their interconnections.

During the data acquisition stage, we collected images from a variety of representative application scenarios, covering a wide range of sample types. The image data originates from multiple mainstream characterization techniques, involving commonly used instruments and diverse imaging modes. This results in a highly heterogeneous image dataset in terms of both material types and imaging modalities.

On this basis, a high-quality annotation system has been established. This system integrates a multi-stage process involving model-based pre-labeling, manual correction, and expert validation, ensuring the accuracy, consistency, and reusability of annotations. Such a robust labeling framework provides a solid foundation for the effective training of deep learning models.

Leveraging these data resources, we trained a microscopy image recognition model with strong generalization capabilities. This model is specifically optimized for the characteristics of microscopy images and is capable of precise segmentation and identification of various target objects. Enabled by this model, the system can automatically perform a wide range of analytical tasks, including pore density analysis, diameter distribution, grain size grading, phase identification, defect detection, length measurement, and sphericity evaluation.

These recognition and analysis modules are broadly applicable in industrial and research domains such as energy materials, catalysts, additive manufacturing, and metal alloys, significantly enhancing the efficiency and objectivity of image analysis.

To facilitate the practical deployment of these research outcomes, we further developed an intelligent analysis platform. This platform supports single-click image processing, batch analysis, high-precision recognition, deep data mining, and visualization. It can automatically generate comprehensive and richly formatted analytical reports, thereby offering a powerful and user-friendly solution for automated and intelligent analysis of microscopy images in both scientific research and industrial applications.

%% file: Sections/03-Data_Engine.tex
\section{Dataset and Data Engine}
\subsection{Data Engine}

In order to support robust instance segmentation model training on high‑density electron microscopy (EM) images, we developed a web‑based data engine that integrates data management, annotation, and quality control. Unlike natural image segmentation—where there are numerous large-scale publicly available datasets—EM image datasets suffer from severe scarcity due to the rarity of instrumentation, proprietary restrictions, and privacy concerns. Normally, annotation tools (e.g. LabelMe (\cite{wada2018labelme}), Label Studio (\cite{LabelStudio})) fail to scale to our high‑density scenarios—2,000 × 2,000 images often contain more than 4,000 individual instances, resulting in unacceptable performance degradation and software instability. To overcome these challenges, we designed and implemented an online platform tailored to our domain, complete with specialized annotation widgets, a multi‑stage review pipeline, and iterative model–human collaboration to improve both data scalability and model performance.

The platform delivers smooth pan‑zoom navigation and region‑of‑interest rendering even when tens of thousands of polygons are present. Crucially, we integrated a magnetic‑lasso tool with real‑time, Sobel‑based edge snapping—allowing annotators to trace complex, low‑contrast boundaries quickly and accurately. Annotators may further adjust snapping sensitivity to suit variations in image contrast and specimen heterogeneity.

Quality control is enforced by a multi‑stage review pipeline: each image is first checked by two independent peer annotators, who flag missing or misaligned masks via inline comment threads, and then undergoes a final audit by an expert in materials science. All comments are anchored to specific vertices or instances, and all edits and review decisions are versioned to ensure traceability and allow rollback if discrepancies emerge.

To bootstrap the dataset and then scale via model‑assisted annotation, we adopted a two‑stage iterative approach. In Stage I, domain experts annotate a curated, low‑density subset of EM images to produce high‑accuracy masks. These gold‑standard annotations train an initial instance segmentation model, which then generates pseudo‑labels on the remaining unannotated data. In Stage II, a larger pool of trained volunteers refines the pseudo‑labels using our annotation tools, and corrected masks are validated through the same quality control pipeline before inclusion. Successive cycles of model retraining and human refinement have expanded our dataset to 5,091 images.

\subsection{Data Generation}
Even though comprehensive annotation and review mechanisms have been established, our collected data exhibits a long-tail distribution problem, which makes building truly multi-scenario, large-scale datasets challenging. To address this issue, we leverage large-scale diffusion models to synthesize simulated images, effectively alleviating the long-tail distribution while increasing our total data volume for large-scale pretraining. Additionally, the emergence of LoRA (\cite{hu2022lora}) fine-tuning techniques has dramatically reduced training costs, enabling customization of specific objects with remarkably small datasets (fewer than 20 samples), making efficient adaptation possible even with limited examples.

\textbf{Efficient domain adaptation and manageable computational resources.} By fine-tuning a large-scale diffusion model, such as SDXL (\cite{podell2023sdxl}), using only 3-5 images paired with a text prompt containing a unique word (e.g., ``A [V] particles'') (\cite{ruiz2023dreambooth}), we achieve efficient adaptation to new scenarios. This method ensures that data generation remains within manageable computational resources while retaining the model's robust capabilities acquired during large-scale pre-training.

\textbf{Controlled Generation for paired dataset.} LoRA-based fine-tuning solutions enable us to quickly learn the visual characteristics of specific materials on small datasets and produce a large number of similar samples. However, within our data engine, only image-mask pairs are meaningful for downstream training. Therefore, to generate paired data, we divide the sample augmentation approach into two phases. In phase one, Image-LoRA is used to perform high-quality fitting of visual concepts of specific materials, while Mask-LoRA is utilized to fit masks of instances of different materials. In subsequent stages, Mask-LoRA and heuristic Python scripts are employed to randomly generate instance segmentation masks in bulk. In phase two, ControlNet (\cite{zhang2023adding}) is introduced, where rendered instance segmentation masks are used as conditional inputs to the diffusion model, with their corresponding original images serving as target outputs. Through training on paired data, the ControlNet enhanced by Image-LoRA can utilize the generated masks from the aforementioned to produce corresponding synthetic images.

\begin{figure}[tp]
    \centering
    \includegraphics[width=\linewidth]{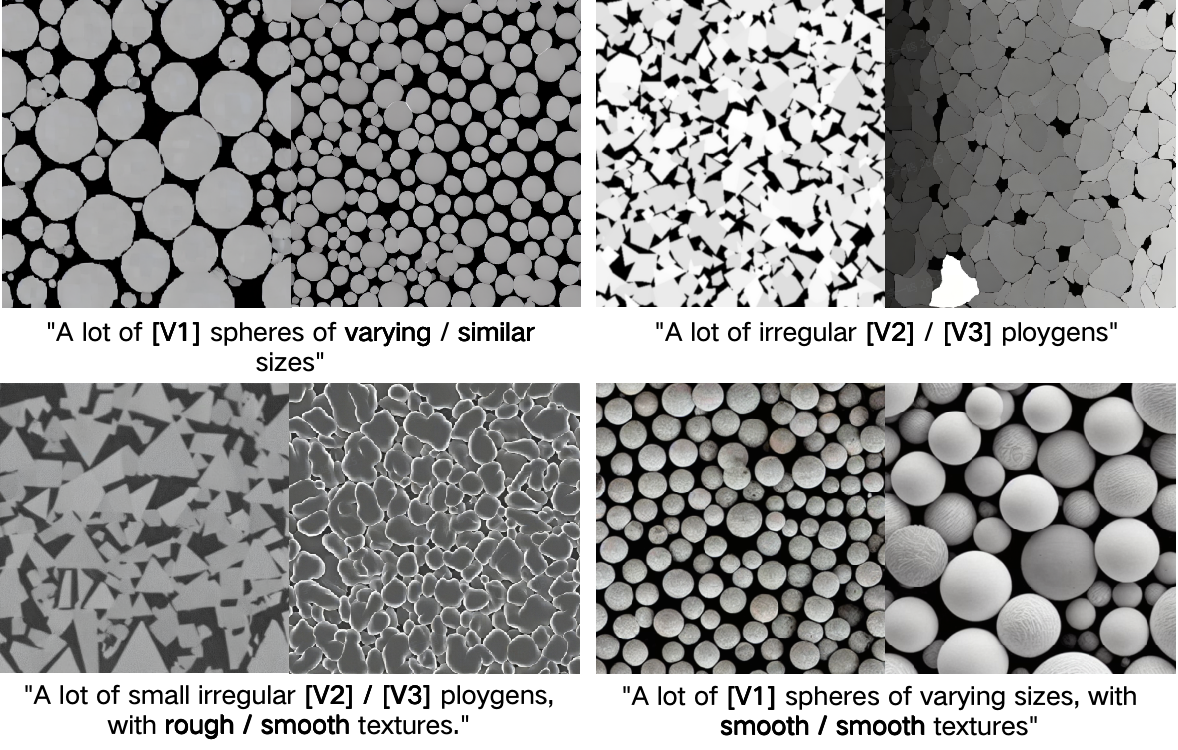}
    \caption{Examples of synthetic data generated by latent diffusion model under different textual prompts.}
    \label{fig:gen}
\end{figure}

The example patterns shown in Figure~\ref{fig:gen} were synthesized using a large‐scale, pre‐trained latent diffusion model guided by simple textual prompts. Specifically, each prompt takes the form “A lot of [V1] spheres of varying sizes,” “A lot of irregular [V2] polygons,” etc., where the placeholder [V] is a unique keyword to enable efficient fine‐tuning to novel scenarios. As can be seen, our approach affords precise control over the generation of granular structures across different scenes, while independently modulating particle shape (e.g., spheres vs. polygons), size distributions (varying vs. similar), and surface texture (rough vs. smooth). These results underscore the flexibility of our method for generating tailored particulate media in simulation tasks.

\subsection{Dataset}
We assembled a large-scale instance segmentation dataset of 5,091 EM images, containing about 3 million annotated instances. We refer to this dataset as UniEM-3M, which has been released and described in detail in our work (\cite{uniem3m}). The dataset was split into a training set of 4128 images and a test set of 963 images. These images were sourced from three primary channels: (1) high-quality, original EMs from academic collaborations, (2) automated web crawling, and (3) the annotation of approximately 500 challenging images from the public dataset (\cite{aversa2018first}).

The images exhibit extensive diversity in both sample origins and imaging modalities, which enhances model generalization and ensures adaptability across varied real-world scenarios. Data were acquired using multiple microscopy techniques—scanning electron microscopy (SEM), transmission electron microscopy (TEM), scanning transmission electron microscopy (STEM), optical microscopy (OM), and focused ion beam (FIB) microscopy—and include back-scattered electron (BSE), secondary electron (SE), and electron back-scatter diffraction (EBSD) modes. Sample materials span biological cells, metallic alloys, ceramics, composites, nanomaterials, semiconductor devices, battery materials, and magnetic materials, covering both rigid inorganic structures and soft biological tissues. Morphological diversity is furthered by images of nanowires, nanoparticles, powders, thin films, and fibers, as well as microstructures such as cross-sectional particles, three-dimensional agglomerates, secondary particles, and pores. These objects vary widely in size (nanometer to micrometer scale), shape (regular to highly irregular), density, and overlap, creating complex, realistic scenarios for evaluating the performance and generalization capabilities of instance segmentation models.

\begin{figure}[t]
    \centering
    \includegraphics[width=\linewidth]{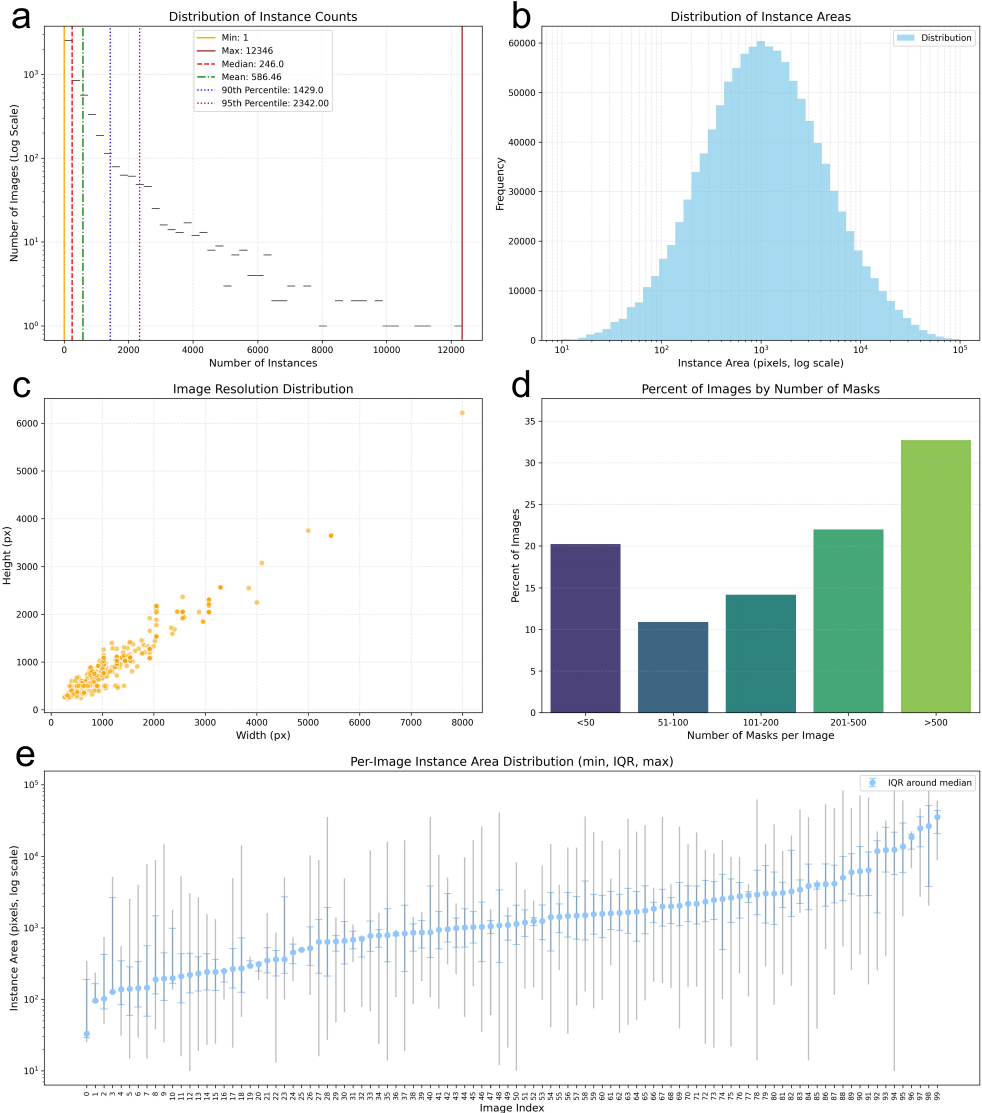}
    \caption{The statistics of our dataset. \textbf{a}, Distribution of instance counts per images (log scale). \textbf{b}, Distribution of instance areas (pixels, log scale). \textbf{c}, Distribution of image resolution. \textbf{d}, Percentage of images by number of masks. \textbf{e}, Per‐image instance area range (min, IQR, max, log scale)}
    \label{fig:data}
\end{figure}

We performed detailed statistical analyses on five key properties of our dataset in Figure~\ref{fig:data}. In Figure 3a, the distribution of instance counts per image is shown on a logarithmic scale: although the median image contains 246 instances (red dashed line), the mean is substantially higher ($\approx$586, green dash–dot), and the distribution exhibits a heavy right tail (90th percentile at 1429, 95th percentile at 2342; blue and purple dotted lines), with a few images containing thousands of instances. Figure 3b displays a log‐scale histogram of individual mask areas: most object masks occupy between $10^2$ and $10^4$ pixels, but both very small and very large instances are present, yielding a positively skewed area distribution. In Figure 3c, a scatter plot of image widths versus heights reveals that image resolutions span a broad range (from roughly 200×200 px up to over 8,000 × 6,000 px), with a dense cluster around 400–1,500 px in each dimension. Then, Figure 3d shows the proportion of images by mask‐count bins: approximately 20 \% of images contain fewer than 10 masks, 11 \% contain 51–100, 14 \% contain 101–200, 22 \% contain 201–500, and 33 \% contain more than 500 masks.  Finally, Figure 3e illustrates the per‐image variation in object instance scale across a representative subset of our dataset. To construct it, we first retained only those images containing more than 200 annotated instances; from this pool we then randomly sampled 100 images. For each selected image we computed the pixel area of every instance and plotted, on a logarithmic scale, the minimum instance area, the interquartile range around the median (blue bar), and the maximum area. The wide vertical spans for most images demonstrate that object scales vary by several orders of magnitude both within individual images and across the dataset as a whole, highlighting the challenge of learning scale‐robust detectors. Collectively, these panels highlight pronounced imbalance both in the number and size of instances as well as in image resolutions.

%% file: Sections/04-Model.tex
\section{Model}

\subsection{Overall Workflow}

\begin{figure}[tp]
    \centering
    \includegraphics[width=0.8\linewidth]{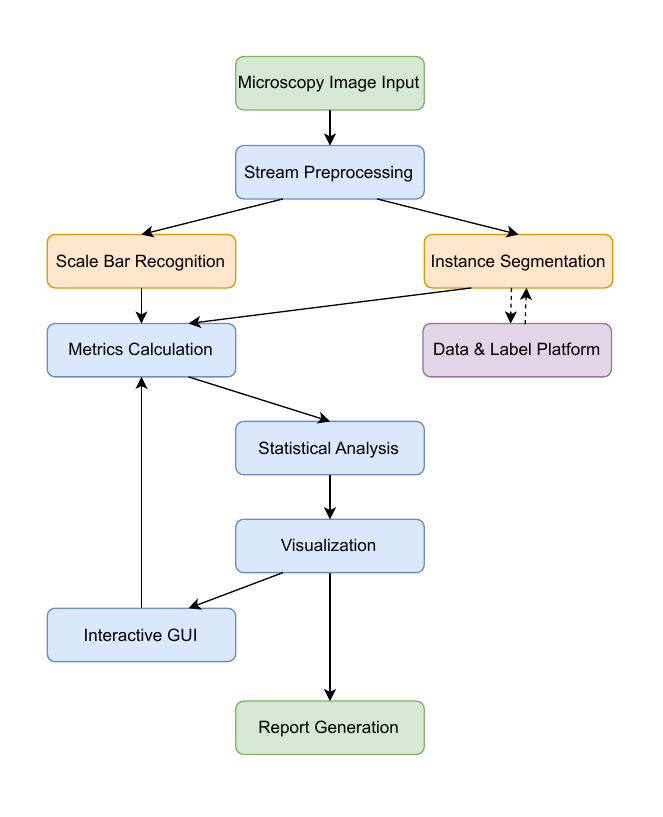}
    \caption{The overall workflow of Uni-AIMS. The results of scale bar detection and instance segmentation are combined to produce statistical analysis. If algorithms give incorrect results, users can correct them using visualization and interactive tools, and recalculate statistics.}
    \label{fig:algo_workflow}
\end{figure}

We designed a comprehensive microscopy image processing and analysis pipeline capable of generating detailed analytical reports from raw images. Figure~\ref{fig:algo_workflow} illustrates the complete pipeline. 

Our workflow begins with batched microscopy image input, which is followed by image preprocessing before branching into parallel processes: scale bar recognition and instance segmentation. 
The instance segmentation module also enables AI-assisted labeling on our own data annotation platform, which helps to refine the quality of training data and further improves algorithm performance. Results from both algorithms feed into metrics calculation, which leads to statistical analysis of the extracted features. 
The analysis results are visualized and presented through both an interactive GUI (which allows human feedback or correction to recompute metrics) and automated report generation, discussed in Section~\ref{sec:platform}. 

In this section, we will further discuss our model and algorithm used in scale bar recognition and instance segmentation.

\subsection{Instance Segmentation}

Microscopy images of particles or cells present unique challenges: densely packed structures, diverse morphologies, and blurred boundaries. Traditional deep learning methods often struggle with these characteristics, particularly when needing to simultaneously process a large number of multi-scale targets.

\subsubsection{Flow-based Segmentation}

Instance segmentation fundamentally addresses the challenge of effectively separating individual instances. Detection-based methods like Mask R-CNN (\cite{he2017mask}) employ a two-stage process of detection followed by segmentation, offering intuitive implementation and clear instance differentiation, but suffering performance degradation with densely packed or heavily overlapping objects. In contrast, one-stage methods like YOLACT (\cite{bolya2019yolact}) prioritize speed by producing shared prototype masks, trading off some mask quality. Center-based methods such as CenterMask (\cite{Wang_2020_CVPR}) and SOLO (\cite{wang2020solo}) construct masks starting from instance centers, effectively handling convex objects in dense scenarios but performing poorly with irregular shapes. Universal segmentation frameworks such as Mask2Former (~\cite{cheng2021mask2former}) have been proposed, which leverage a transformer-based decoder with masked attention to unify semantic, instance, and panoptic segmentation. While highly general and flexible, these methods may struggle to maintain fine-grained mask quality in microscopy images, particularly under extreme density or scale variation. StarDist (\cite{schmidt2018}) and its extension CPP-Net (~\cite{chen2023cpp}), although often grouped into center-based approaches, actually model each object as a star-convex polygon via radial distances from a predicted center point, which makes it particularly well-suited for convex cell nuclei but less effective on irregularly shaped structures. Flow-based methods like HoverNet (\cite{graham2019hover}), Cellpose (\cite{stringer2021cellpose}) and CellViT (\cite{horst2024cellvit}) predict directional fields from pixels to instance centers, efficiently managing complex shapes in densely packed arrangements—particularly suitable for microscopy image analysis—though requiring computationally intensive post-processing and highly accurate flow prediction.

Our method uses flow field prediction as the core of instance segmentation, primarily based on its ability to effectively address the following microscopy image characteristics:
\begin{itemize}
    \item Densely packed targets with minimal separation
    \item Significant variations in target size and shape
    \item Complex boundary structures with gradient transitions rather than clear edges
    \item Large number of targets in a single image, creating computational challenges
\end{itemize}

The flow-based representation enables the model to learn directional fields from geometric boundaries to centers, guiding the simultaneous separation of individual instances even in highly dense environments. This characteristic makes it particularly suitable for microscopy image analysis tasks.

\begin{table*}[htbp]
\centering
\caption{Baseline performance comparison of various instance segmentation methods on sparse and dense subdatasets.}
\begin{tabular}{llcccccc}
\toprule
\multirow{2}{*}{Class} &\multirow{2}{*}{Method} & \multirow{2}{*}{\centering Backbone} & \multirow{2}{*}{\centering Params} &\multicolumn{2}{c}{Sparse} & \multicolumn{2}{c}{Dense} \\
        &    &  &  & $mAP@0.5$ & $PQ@0.5$ & $mAP@0.5$ & $PQ@0.5$  \\
\midrule
\multirow{3}{*}{Anchor-based}        & Mask R-CNN         &  ResNeXt101 & 101M & 0.542 & 0.565 & - & - \\
                               & Cascade R-CNN    &  ResNeXt101 & 135M & 0.540 & 0.587 & - & - \\
                               & HTC              &  ResNeXt101 & 137M & 0.345 & 0.461 & - & - \\
\midrule
\multirow{4}{*}{Anchor-free}   
                               & YOLACT         &  ResNet101 & 54M & 0.307 & 0.293 & - & - \\
                               & Mask2Former & Swin-base & 107M & 0.204 & 0.064 & - & - \\
                               & StarDist           &    SAM-ViT-base    &  140M    & 0.236 & 0.230 & 0.487 & 0.438 \\
                               & CPP-Net        &     SAM-ViT-base    &   140M   & 0.196 & 0.187 & 0.378 & 0.352 \\
                               
\midrule
\multirow{3}{*}{2D Field}    
                              & CellViT         &    SAM-ViT-base       &   146M   & 0.589 & 0.573 & 0.701 & 0.604 \\
                              & Cellpose-SAM    & SAM-ViT-large & 304M & 0.605 & 0.633 & 0.760 & 0.700 \\
                              & Uni-AIMS         & SAM-ViT-base  &  93M & \textbf{0.824} & \textbf{0.720} & \textbf{0.787} & \textbf{0.703} \\
\bottomrule
\end{tabular}%
\label{tab:seg_AP}
\end{table*}

\begin{figure*}[ht]
    \centering
    \includegraphics[width=0.95\linewidth]{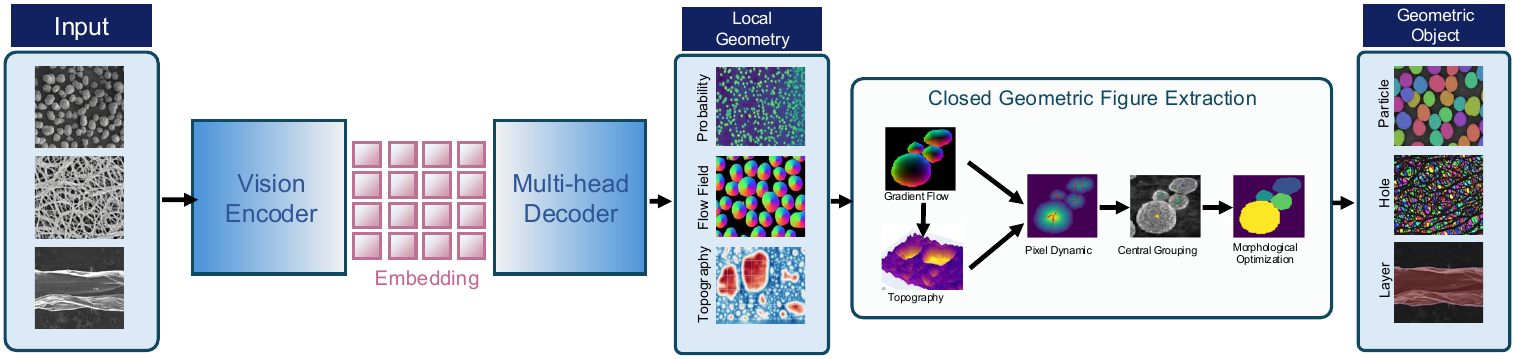}
    \caption{The overview of our instance segmentation algorithm. The input images are encoded by a common vision encoder to patched image embeddings. The embeddings are optionally enhanced with visual and text prompt tokens and sent into a multi-head decoder to get geometry information. After post-processing of pixel dynamics, well-separated geometric objects are extracted.}
    \label{fig:flowmodel}
\end{figure*}

Our flow-based segmentation algorithm is illustrated in Figure~\ref{fig:flowmodel}. It integrates local geometric feature extraction with global structure analysis, forming a comprehensive processing framework. 
The method first extracts local morphological and geometric information from the raw image, including boundary gradient fields and foreground object skeletons. 
These low-level features accurately capture the morphological structure of objects, laying the foundation for subsequent analysis. 
For closed geometric figures, we apply modified Euler integration to the predicted flow fields, creating pixel dynamics models that guide all foreground pixels toward specific convergence centers. 
By integrating local geometric information with global flow dynamics, our algorithm achieves precise segmentation of targets with different morphologies and scales, providing a reliable foundation for subsequent quantitative analysis and feature extraction. 
This method is not only applicable to closed morphological cells and particles but also effectively handles open-structured fiber networks, demonstrating broad application potential.

\subsubsection{Implementation}

Our algorithm is independent of specific network architecture, making it possible to switch between neural network models of varying computational complexity based on the specific application scenario. For applications with limited computational resources and no need for interactive prompts, we employ a U-Net structure (\cite{ronneberger2015u}) with 2D convolution and skip connections as the core architecture for encoders and decoders. When sufficient data is available, incorporating larger backbone networks, such as ResNet (\cite{he2016deep}), ResNeXt (\cite{xie2017aggregated}), ViT (\cite{dosovitskiy2020image}) or Swin (\cite{liu2021swin}) can further enhance model performance, achieving superior analytical results.

In similar algorithms, post-processing based on pixel dynamics often constitutes the main performance bottleneck. To address this issue, we implemented the pixel dynamics processing workflow using CUDA, significantly improving the system's inference throughput. Currently, our optimization solution can complete comprehensive processing of a single 8K$\times$8K ultra-high-resolution image in approximately 80 seconds on a single Nvidia 4090D GPU, substantially increasing practical application efficiency.

\subsubsection{Result}
We adopt standard metrics for instance segmentation: mean Average Precision (mAP) for mask accuracy, and Panoptic Quality (PQ) (~\cite{kirillov2019panoptic}) at 0.5 for combined segmentation and recognition quality. Following StarDist (~\cite{schmidt2018}), we use AP calculated by $\text{AP} = \frac{|TP|}{|TP| + |FP| + |FN|}$ \footnote{In this paper, AP follows the convention in bioimage segmentation~\cite{stringer2021cellpose,graham2019hover,schmidt2018}, which differs from the COCO-style AP but is widely adopted in microscopy instance segmentation.} at an IoU threshold $T$, where the predicted masks matching true masks over $T$ are regarded as TP. Mean AP is the average AP of all images. 

Anchor-based methods (e.g., Mask R-CNN, Cascade Mask R-CNN) and recent frameworks such as YOLACT and Mask2Former often suffer from excessive computational cost and degraded accuracy under ultra-dense microscopy scenes. 
To better reflect these performance discrepancies, we divide our dataset into two subsets: a \textit{sparse} set (images with instances fewer than 100) and a \textit{dense} set (more than 100 instances). 
In contrast, bioimage-oriented methods like Cellpose and HoverNet are generally more robust to both regimes, with memory usage largely independent of the number of instances. 
All compared methods in Table~\ref{tab:seg_AP} were carefully tuned and fine-tuned on our dataset to ensure a fair comparison. These experiments demonstrate the superior capability of flow-based methods, particularly Uni-AIMS, in handling high-density microstructures, likely attributable to their effective use of 2D fields for instance separation in crowded scenes. 

\subsection{Scale Bar Recognition}

Scale bars in microscopy images are crucial for quantitative image analysis, providing the necessary reference to convert between pixel dimensions and real-world physical sizes. Scale bar recognition typically involves two key steps: detecting and localizing the scale bar to obtain its pixel length, and extracting the corresponding physical size information (such as numeric value and unit).

Existing methods primarily include:
\begin{itemize}
    \item \textbf{Manual Measurement}: Time-consuming, subjective, and difficult to automate.
    \item \textbf{Threshold-based Image Segmentation}: Binarizing the image using fixed thresholds to separate the scale bar from the background. This approach performs well for high-contrast and simple backgrounds but degrades significantly under complex or noisy conditions.
    \item \textbf{Edge Detection Algorithms}: Employing operators such as Canny (\cite{canny1986computational}) or Sobel (\cite{danielsson1990generalized}) to detect image edges for locating scale bars. However, these methods are prone to noise, often resulting in false detections or missed detections.
    \item \textbf{Template Matching Techniques}: Matching pre-defined scale bar templates to the image, which is effective for standardized scale bar styles but struggles with diverse or deformed scale bars.
\end{itemize}

Many existing automated methods (\cite{mukaddem2019imagedataextractor}, \cite{kim2020machine}) are designed for specific equipment manufacturers or fixed styles of scale bars, lacking generalizability and thus failing to address the challenges posed by the highly variable scale bar styles and complex backgrounds in electron microscopy images. Therefore, we propose a robust, efficient, and adaptive scale bar recognition model tailored for electron microscopy images, aiming to achieve accurate detection and physical size extraction under diverse imaging conditions.

\subsubsection{Data Generation}

Our dataset covers a wide range of scale bar types. To expand the training set, we extracted 366 background images without scale bars from collected microscopy images. Based on common scale bar types, information panel styles, fonts, and manufacturer logos, we performed random combinations and augmentations to synthesize approximately 50,000 labeled images.

The random variations included font styles, colors, positioning, measurement units, and scale bar lengths, aiming to significantly enhance the model’s generalization capability.

\subsubsection{Scale Bar and Text Detection}

We adopted the YOLOv8m model (\cite{yolov8_ultralytics}) for the joint detection of scale bars and associated text regions. During training, moderate data augmentation techniques such as horizontal flipping, color jittering, and small-angle rotations were applied. The input image size was set to 1280 pixels, and a 1.5$\times$ down-sampling of the image width was performed to accommodate the typical aspect ratios of scale bars.

For improved inference speed and efficiency, the model was exported to ONNX format and accelerated with FP16 precision.

During inference, the model detects all candidate scale bars and text regions. The detected scale bars are then ranked based on confidence scores, and the one with the highest confidence is selected as the final scale bar instance.

\subsubsection{Scale Bar Endpoint Localization}

To improve the precision of pixel-length measurements, we performed fine-grained processing on the detected scale bar region. Specifically:
\begin{itemize}
    \item Selecting the image channel with the richest edge information;
    \item Applying local thresholding to generate a high-contrast binary image;
    \item Extracting vertical edge features for structural analysis of the scale bar;
    \item Using a peak detection algorithm (analyzing peak amplitude and sharpness) to accurately localize the endpoints of the scale bar.
\end{itemize}

This method combines the strengths of edge detection, thresholding, and signal analysis, enabling robust and accurate extraction of scale bar endpoints even under noisy or complex imaging conditions, thereby significantly reducing the likelihood of misidentification.

\subsubsection{Text Recognition and Scale Bar Information Matching}

For detected text regions, we applied a confidence threshold (0.15) to filter valid text candidates and used the PaddleOCR engine (\cite{ppocr}) for text recognition.

The recognized text content was parsed using regular expressions (pattern matching) to extract effective measurement values and units (e.g., ``5~$\mu$m'', ``2~nm''). Based on spatial proximity, each recognized text segment was matched to its nearest scale bar.

Finally, by combining the pixel length of the scale bar and the parsed physical size information, automatic conversion from pixel dimensions to real-world measurements was achieved.

\subsubsection{Evaluation}

We constructed a test dataset consisting of 111 real-world electron microscopy images, encompassing scale bars from different manufacturers, various styles, and varying background complexities. The evaluation results are summarized in Table~\ref{tab:scalebar_evaluation_results}.
\begin{table}[h]
    \centering
    \caption{Evaluation results on the microscopy image test set.}
    \label{tab:scalebar_evaluation_results}
    \begin{tabular}{lccc}
        \toprule
        Metric & Unit Accuracy & Value Accuracy & Scale Bar MAE (px) \\
        \midrule
        - & 100\% & 100\% & 0.63 pixels \\
        \bottomrule
    \end{tabular}
\end{table}

Across all test samples, both the unit and value recognition achieved 100\% accuracy, and the average pixel error for scale bar length measurements was only 0.63 pixels, demonstrating the model’s robustness and high precision under diverse imaging conditions.

\begin{figure*}[htbp]
    \centering
    \includegraphics[width=0.99\textwidth]{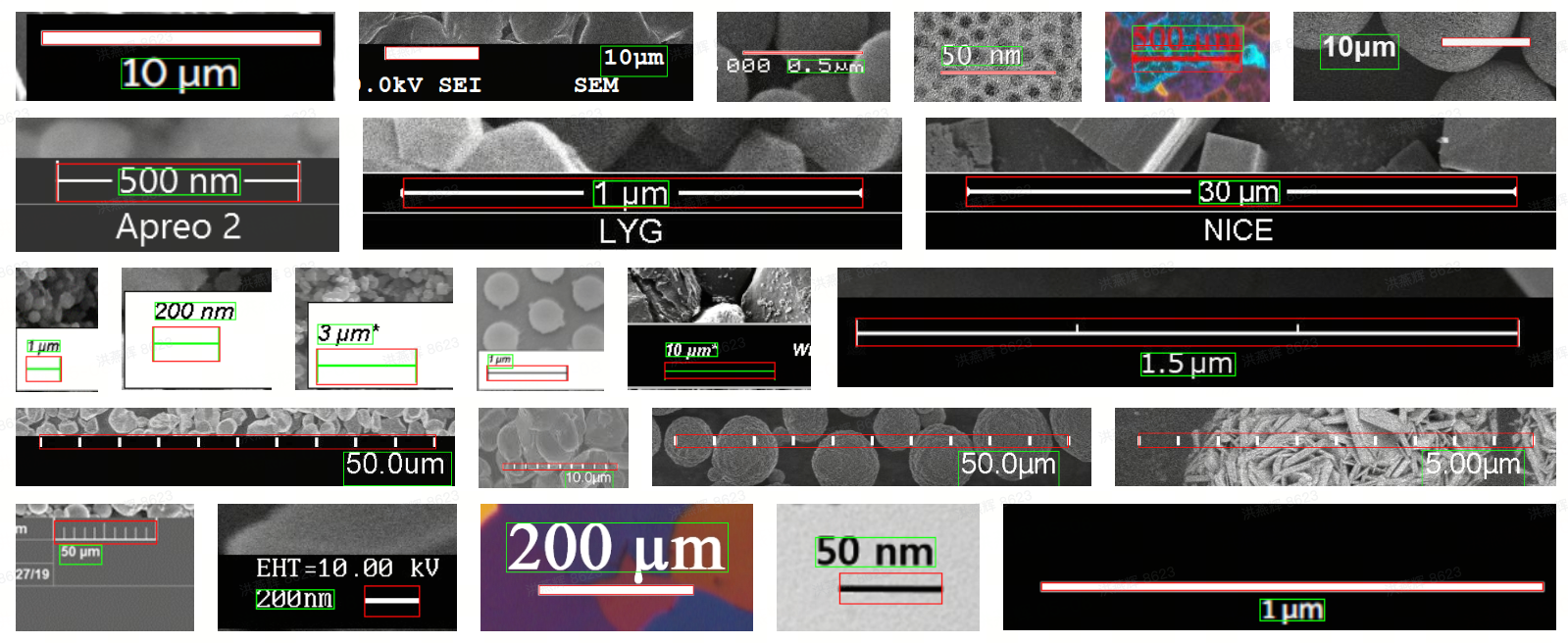}
    \caption{Recognition results of different style scale bars in the test dataset. The red rectangle indicates the location of the scale bar identified by the model. The green rectangle represents the text box corresponding to the scale bar value and unit identified.}
    \label{fig:scalebar_vis}
\end{figure*}

The results of scale bar recognition for some images from test dataset are shown in Fig.~\ref{fig:scalebar_vis}. The red rectangular boxes indicate the positions of the recognized scale bars, while the green rectangular boxes denote the positions of the recognized scale bar text.

\begin{figure}[htbp]
    \centering
    \includegraphics[width=0.49\textwidth]{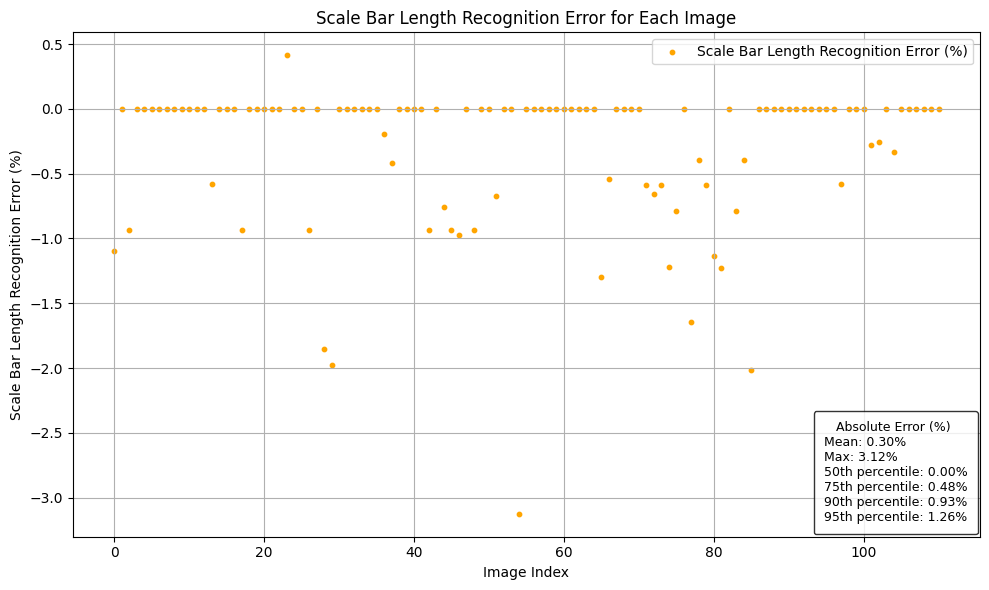}
    \caption{Error between recognized and ground truth scale bar lengths across test dataset.}
    \label{fig:scalebar_length_error}
\end{figure}

Figure \ref{fig:scalebar_length_error} shows the error between the scale bar lengths recognized by our model and the ground truth values across different images. Each point represents the error for an individual image, calculated as (recognized length - ground truth length)/ground truth length.
The text box in the lower right corner shows key statistics of the absolute errors, including mean and percentile values. The results indicate that the majority of errors are very small, with a mean absolute error of 0.3\%. And more than 50\% of the data have an identification length that is exactly correct. Even at higher percentiles (75\%, 90\%, 95\%), the errors remain relatively small, indicating consistent performance across most images.


%% file: Sections/05-Platform.tex
\section{Analysis Platform}
\label{sec:platform}

Based on the results of instance segmentation and scale recognition, we developed Uni-AIMS (\cite{uniaims}) - the intelligent microscopic image analysis platform. This platform performs particle metric calculations, statistical analysis, data visualization, and automated report generation, offering a one-stop intelligent analysis solution for users.

\subsection{Metric Calculation and Statistics}

The platform calculates various key metrics of the recognized particles. These metrics include, but are not limited to, diameter, shape (such as aspect ratio and sphericity), area, and smoothness. Through statistical analysis of these metrics, the platform generates various statistical measures, such as count, maximum, minimum, mean, standard deviation, P10, P50, and P90, for different indicators to comprehensively characterize the distribution and central tendency of the data.

\subsection{Data Visualization}

To intuitively present the recognition and statistical results, the platform integrates robust data visualization capabilities and supports multiple types of charts:
\begin{itemize}
    \item \textbf{Recognition Results}: Visualize the results of model recognition on the image to help users intuitively understand the objects of analysis.
    \item \textbf{Distribution Histograms}: Display the distribution of different metrics, such as the diameter distribution histograms.
    \item \textbf{Scatter Plots}: Illustrate the correlation between different metrics, such as the relationship between diameter and sphericity.
    \item \textbf{Box Plots/Violin Plots}: Compare the distribution differences between different images.
\end{itemize}

These charts not only help users quickly understand the distribution and characteristics of the data but also support customization of styles and parameter adjustments to meet the needs of different users.

\subsection{User Interaction and Reporting}

The platform offers users an efficient and flexible interactive experience through a user-friendly interface, enabling them to perform the following tasks with minimal effort:

\begin{itemize}
    \item \textbf{Result Visualization}: Users can freely select the metrics and chart types to display and adjust chart parameters in real-time (such as the chart colors).
    \item \textbf{Interactive Analysis}: Users can directly filter or further analyze the recognition results through the interface, such as selecting particles within a specific diameter range for detailed analysis.
    \item \textbf{Report Generation}: The platform supports one-click generation of analysis reports, which include detailed statistical information, charts, and summary of the analysis results.
\end{itemize}

\begin{figure*}[htbp]
    \centering

    \begin{subfigure}[b]{0.32\textwidth}
        \centering
        \includegraphics[width=\textwidth]{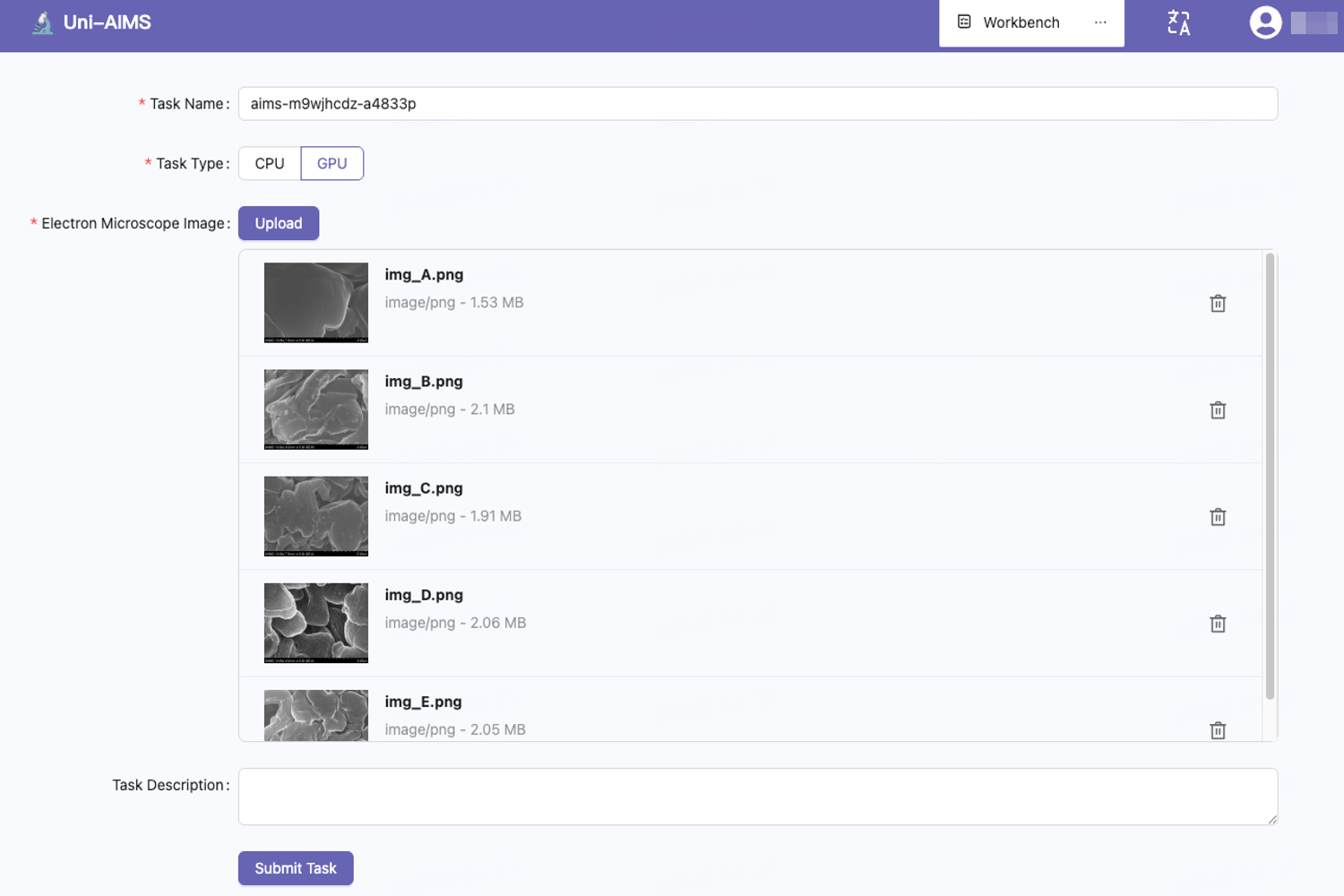}
        \caption{Create Task}
        \label{fig:create_task}
    \end{subfigure}
    \hfill
    \begin{subfigure}[b]{0.32\textwidth}
        \centering
        \includegraphics[width=\textwidth]{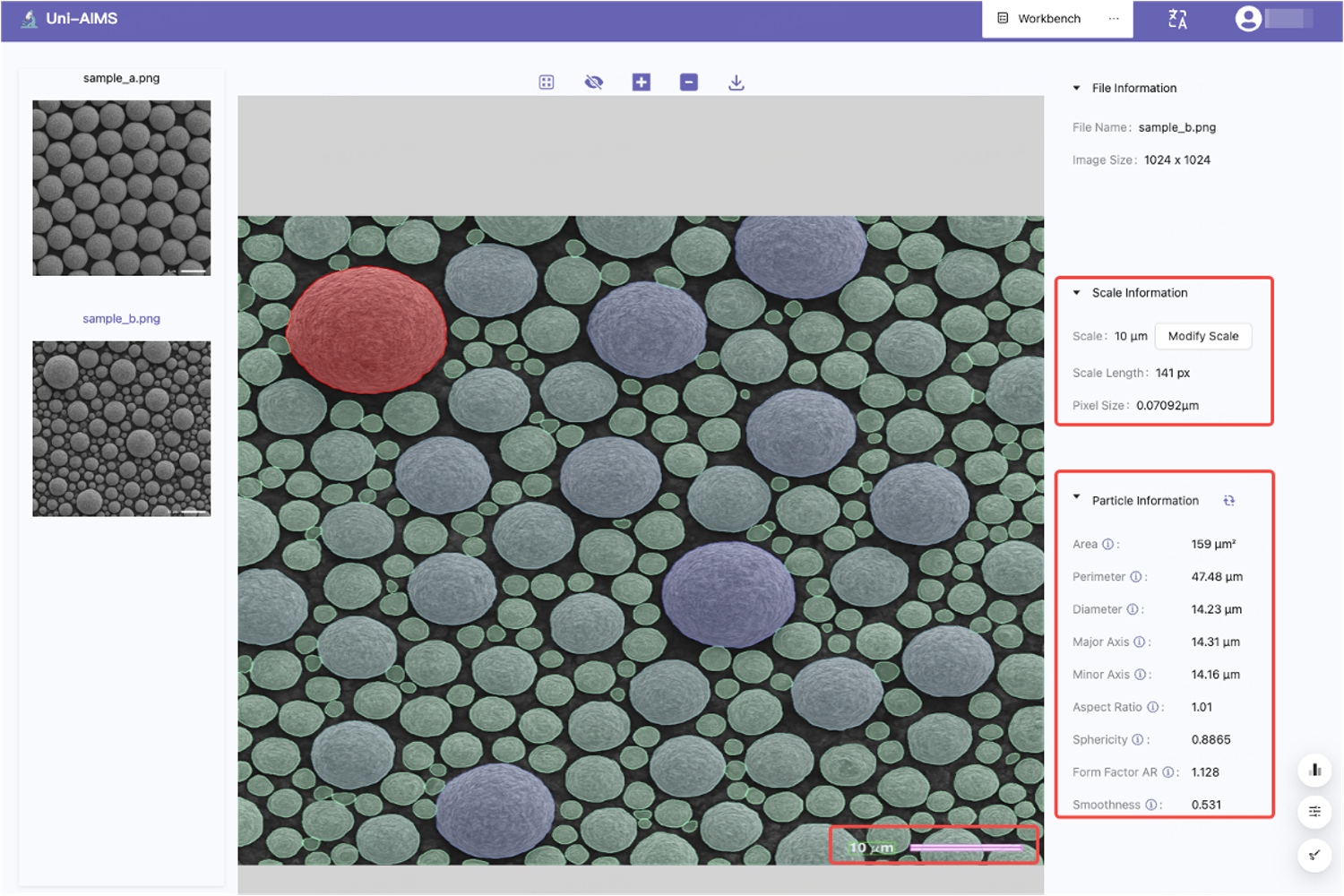}
        \caption{Result Visualization}
        \label{fig:platform_visualize}
    \end{subfigure}
    \hfill
    \begin{subfigure}[b]{0.32\textwidth}
        \centering
        \includegraphics[width=\textwidth]{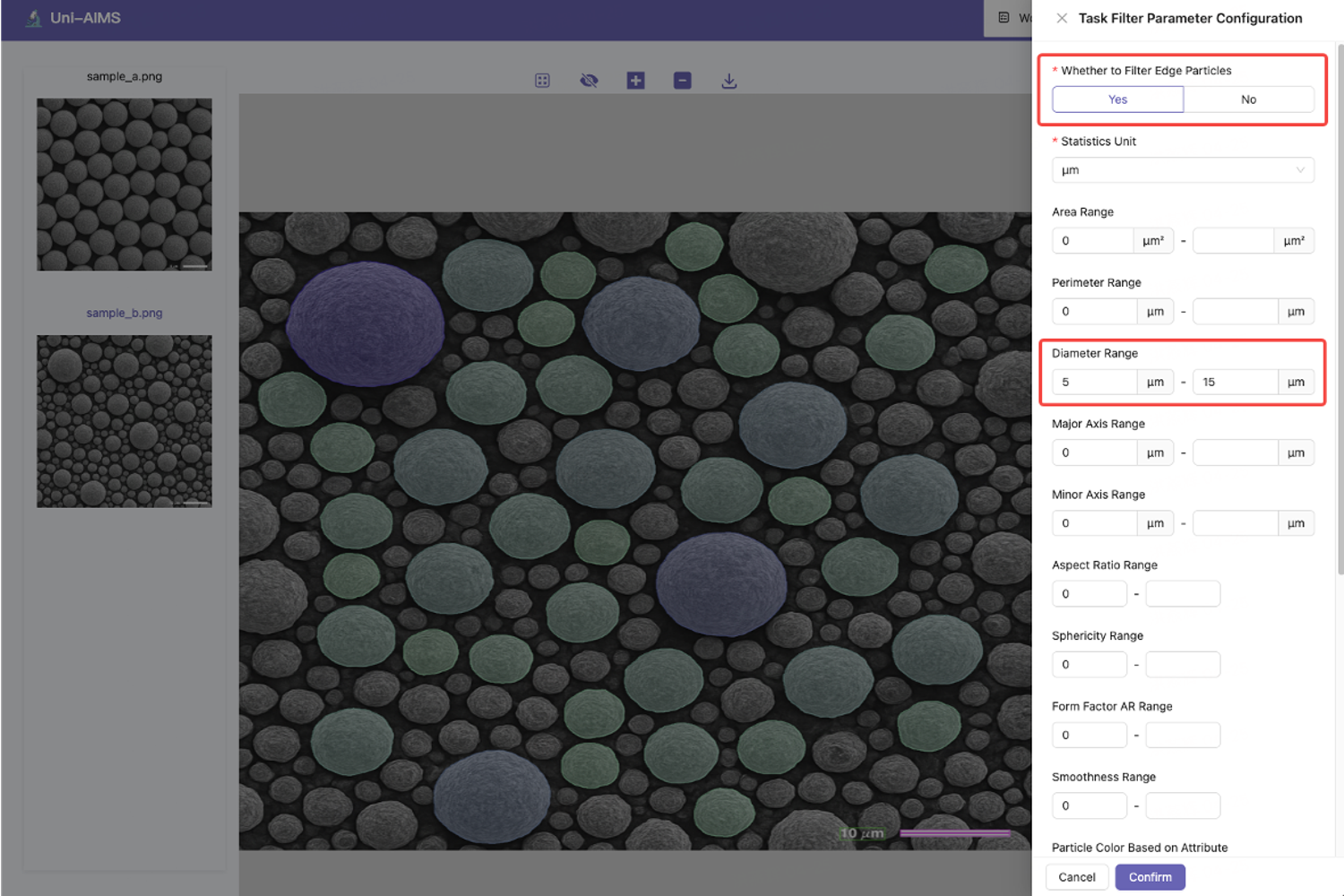}
        \caption{Filter}
        \label{fig:filter}
    \end{subfigure}

    \vspace{0.3cm}

    \begin{subfigure}[b]{0.32\textwidth}
        \centering
        \includegraphics[width=\textwidth]{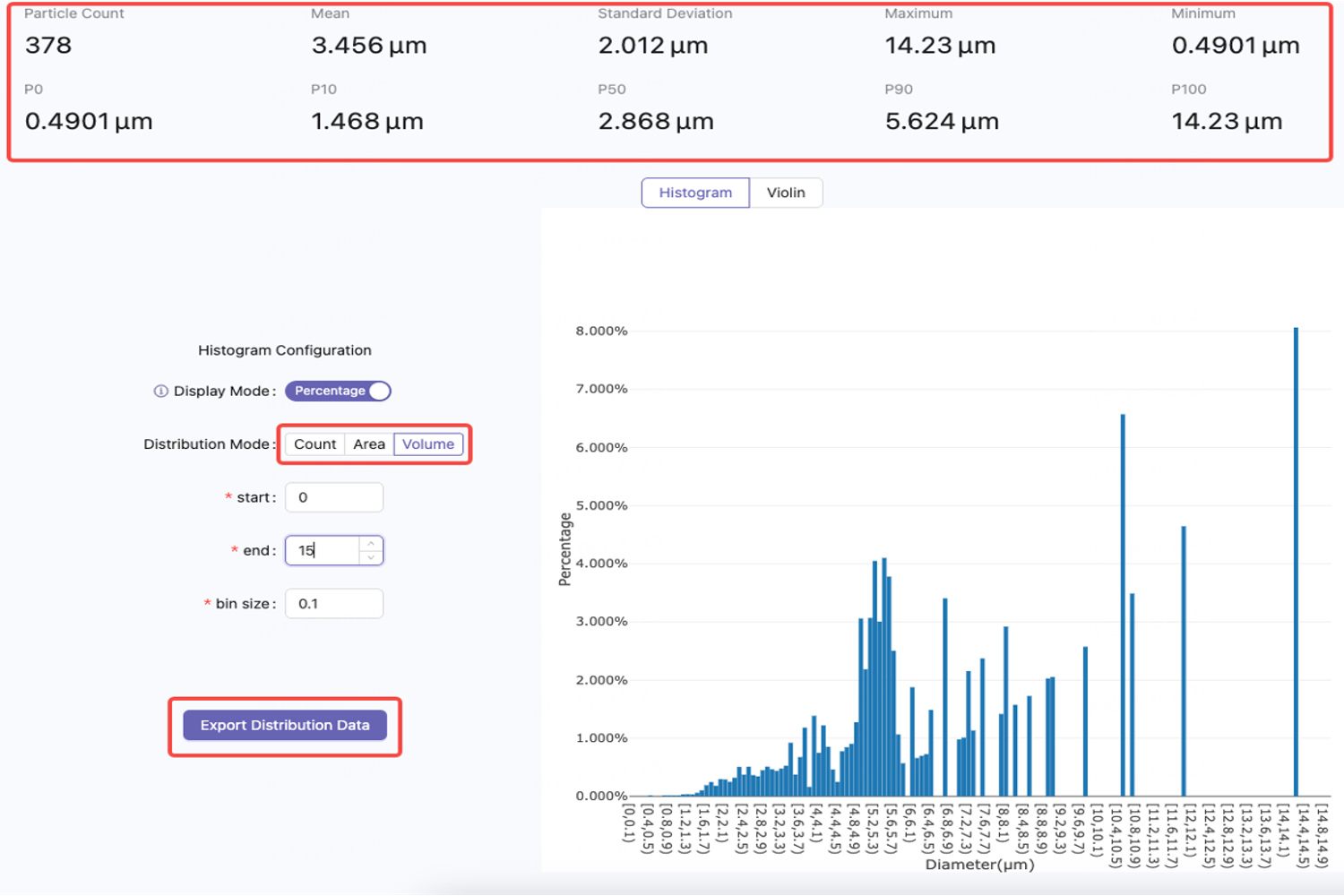}
        \caption{Statistical Information}
        \label{fig:platform_static}
    \end{subfigure}
    \hfill
    \begin{subfigure}[b]{0.32\textwidth}
        \centering
        \includegraphics[width=\textwidth]{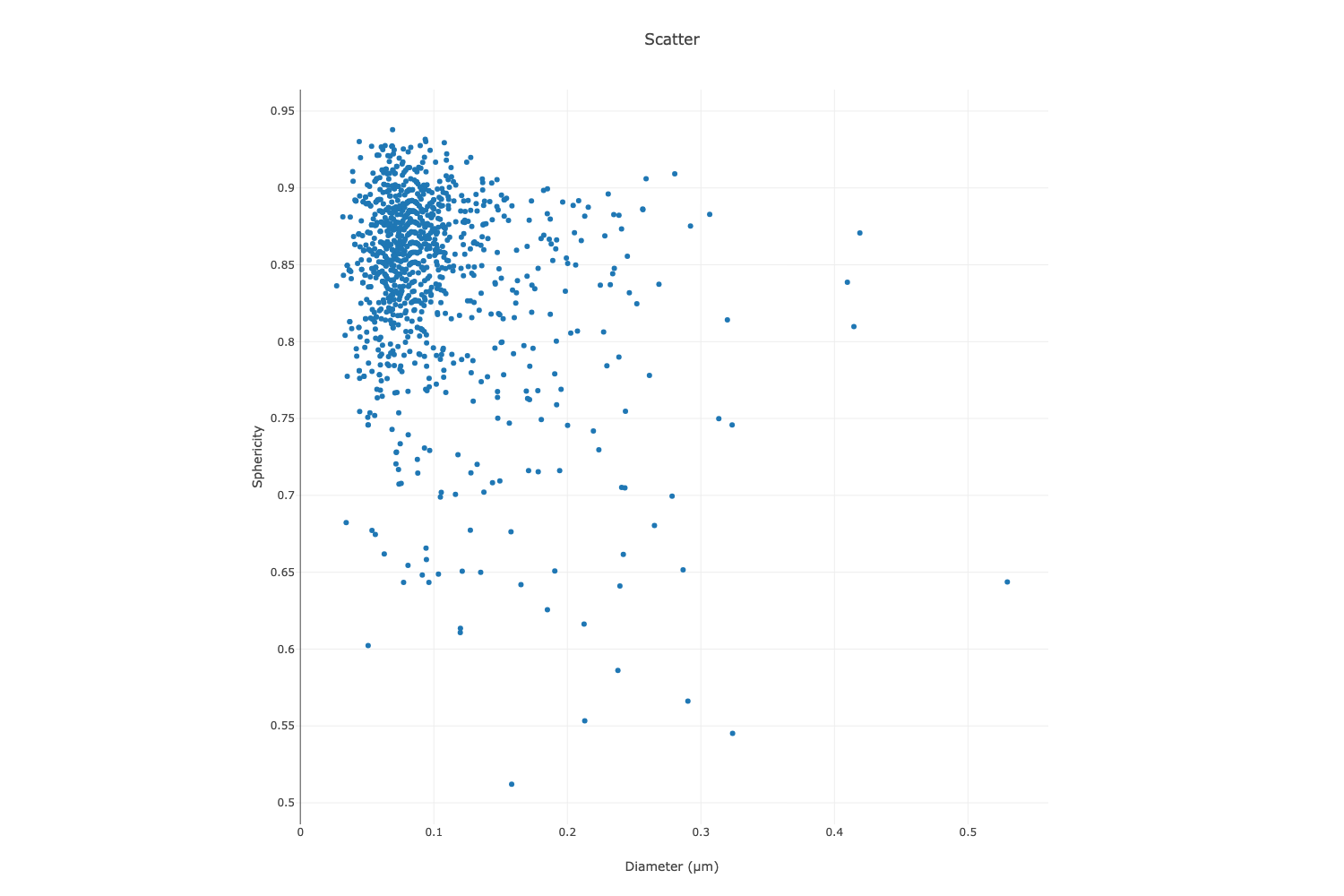}
        \caption{Property Analysis}
        \label{fig:platform_property_analysis}
    \end{subfigure}
    \hfill
    \begin{subfigure}[b]{0.32\textwidth}
        \centering
        \includegraphics[width=\textwidth]{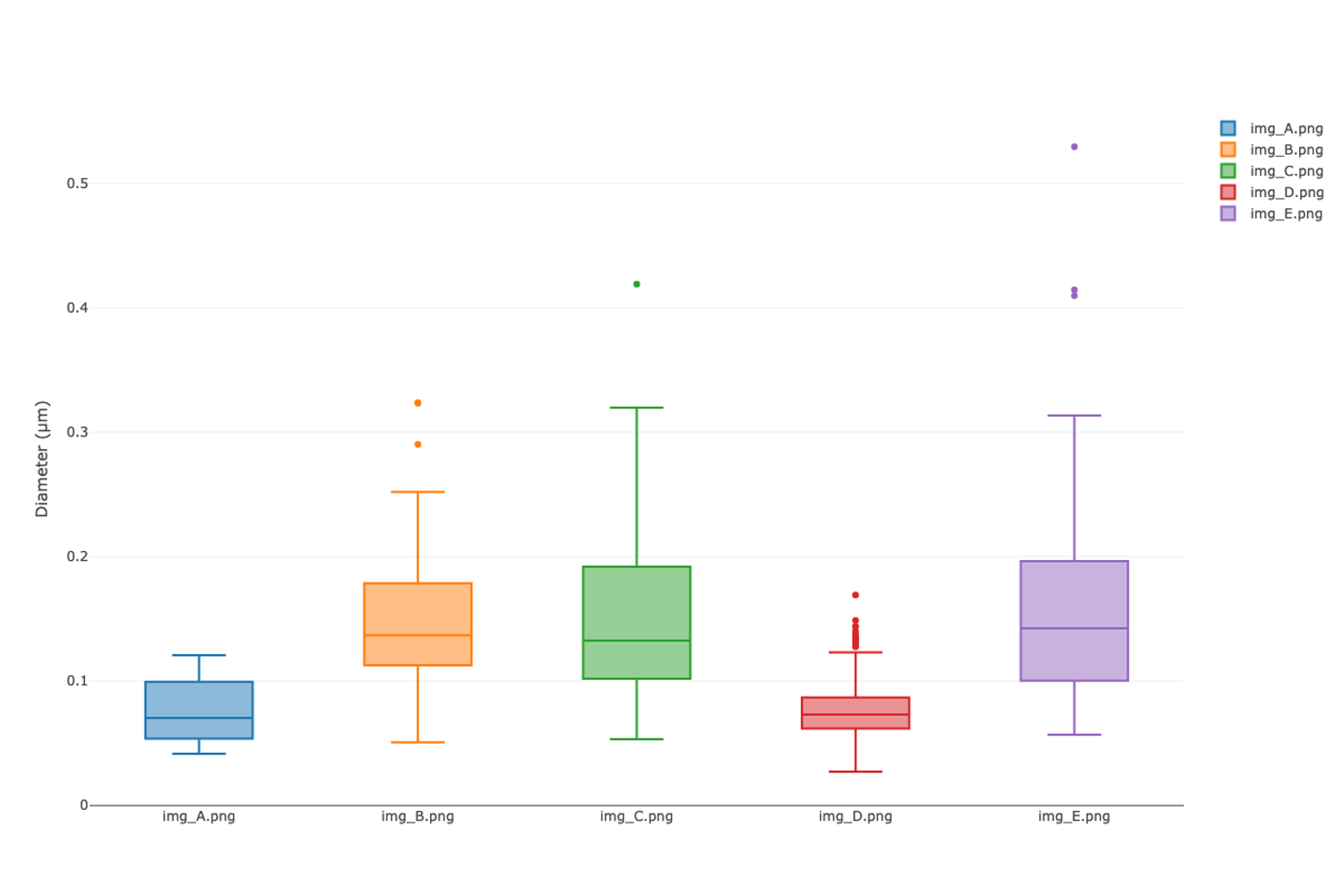}
        \caption{Diameter Comparison of Different Samples}
        \label{fig:platform_comparison}
    \end{subfigure}

    \caption{The screenshots of key functionalities of our microscopy image intelligent analysis platform. It includes: (a) interface for creating analysis tasks; (b) the visualization panel of the recognition result; (c) the filtering function to focus on specific data features; (d) a section providing statistical information; (e) the property analysis feature explores the correlations between two metrics; and (f) the comparison feature for analyzing the diameter of particles across different samples.}
    \label{fig:platform-fig}
\end{figure*}

Figure \ref{fig:platform-fig} illustrates several key operational features of the platform. Figure~\ref{fig:create_task} shows how an analysis task is created on the platform. Users simply upload the images to be analyzed, and the platform automatically performs image processing and recognition. The entire process requires no manual intervention and returns results within a few seconds.

Figure~\ref{fig:platform_visualize} presents the visualization of the recognition results. The central panel is overlaid with masks of the particles detected by the model, along with the identified scale bar and corresponding text. The right panel displays detailed information about the scale bar, as well as a set of metrics of the currently selected particle.

Figure~\ref{fig:filter} demonstrates the interactive capabilities of the platform. Users can filter particles by specifying a diameter range (e.g., 5--15~$\mu$m) and choosing to exclude incomplete particles located at the image edges. After filtering, only particles meeting the criteria are displayed in the central region of the image. These filtered data are then used for subsequent statistical analyses.

Figure~\ref{fig:platform_static} shows the statistical information of particle diameter and the automatically generated histogram. The platform allows users to customize the histogram configuration, including adjusting bin widths and selecting the distribution type (by count, area, or volume), to support diverse analytical needs.

Figure~\ref{fig:platform_property_analysis} displays a scatter plot automatically generated by the platform, depicting the relationship between particle diameter and sphericity. This visualization facilitates the exploration of correlations between particle dimensions and shape characteristics.

Figure~\ref{fig:platform_comparison} illustrates box plots of particle diameter distributions across different image samples, automatically generated by the platform. These plots enable comparison of particle diameter characteristics among different samples.

\subsection{Platform Architecture and Implementation}

The platform employs a modern technical architecture to ensure efficient data processing and a smooth user experience:

\begin{itemize}
    \item \textbf{Front-end}: The interactive interface is built using the React framework (\cite{react}), combined with the Ant Design component library (\cite{antdesign}). This leverages its rich set of components and aesthetically pleasing design style to quickly create high-quality user interfaces. High-performance graphics rendering is achieved using Canvas to meet the needs of complex visualization scenarios.
    
    \item \textbf{Back-end}: FastAPI (\cite{fastapi}) is selected for data processing and interface development, offering high performance and ease of development and debugging, allowing for rapid response to front-end requests. RPC technology (\cite{rpc}) is used to enable efficient communication between different services, facilitating the construction and scalability of distributed systems. The database integrates MongoDB (\cite{mongodb}), a document-oriented database that can flexibly store unstructured and semi-structured data, meeting diverse data storage needs.
    
    \item \textbf{Data Visualization}: D3.js (\cite{d3}) is used for in-depth data analysis and statistical distribution calculations to uncover patterns and characteristics within the data. Plotly.js (\cite{plotly}) is employed to generate various types of professional charts, such as frequency distribution histograms and violin plots, presenting the data in an intuitive and clear manner to users.
\end{itemize}

Through this architectural design, the Uni-AIMS platform provides a one-stop solution for microscopy image analysis, significantly improving analysis efficiency and user experience.

%% file: Sections/06-Application.tex
\section{Application Scenario}

\subsection{Assisting the Analysis of Diameter Distribution of NCM Precursor Particles for Lithium-Ion Batteries}

\subsubsection*{Problem Description}

Lithium batteries are a core power source for new energy vehicles, portable electronics, and energy storage systems due to their high efficiency and environmental friendliness (\cite{armand2008building}). The cathode material is a key component that directly impacts the battery's energy density, cycle life, and safety. Among various cathode materials, nickel-cobalt-manganese (NCM) ternary materials are widely used for their excellent performance (\cite{zhang2024recent}). The preparation and characteristics of NCM precursors are crucial in determining the final cathode performance. Specifically, the particle size distribution of precursors significantly affects the energy density, rate performance, and cycle life of the battery. For example, a narrow particle size distribution can improve crystallinity and electrochemical performance. Smaller particles enhance rate performance due to their higher surface area but may reduce cycle life, while larger particles have the opposite effect (\cite{lee2022co}, \cite{wang2025unveiling}).

Despite the critical importance of particle diameter distribution for battery performance, current methods for measuring and analyzing particle diameter distribution still face many challenges. Traditional methods for measuring particle diameter distribution mainly rely on manual measurement, which is not only inefficient but also prone to subjective influences, resulting in poor accuracy and repeatability of the measurement results. Furthermore, for particles with complex shapes or those that overlap, traditional image processing methods struggle to accurately identify and distinguish them. These issues are particularly prominent in large-scale production, severely affecting the research and development efficiency of battery materials and the quality control of production.

\subsubsection*{Solution}

Based on our model and platform, we are able to automatically identify precursor particles in electron microscopy images. By segmenting each particle in the image and calculating the area of the segmentation mask and the equivalent circular diameter, we can obtain the particle diameter of each particle. We can further statistically analyze the particle diameter distribution and generate a visual report.

\begin{figure}[htbp]
    \centering
    \begin{subfigure}[b]{0.23\textwidth}
        \centering
        \includegraphics[width=\textwidth]{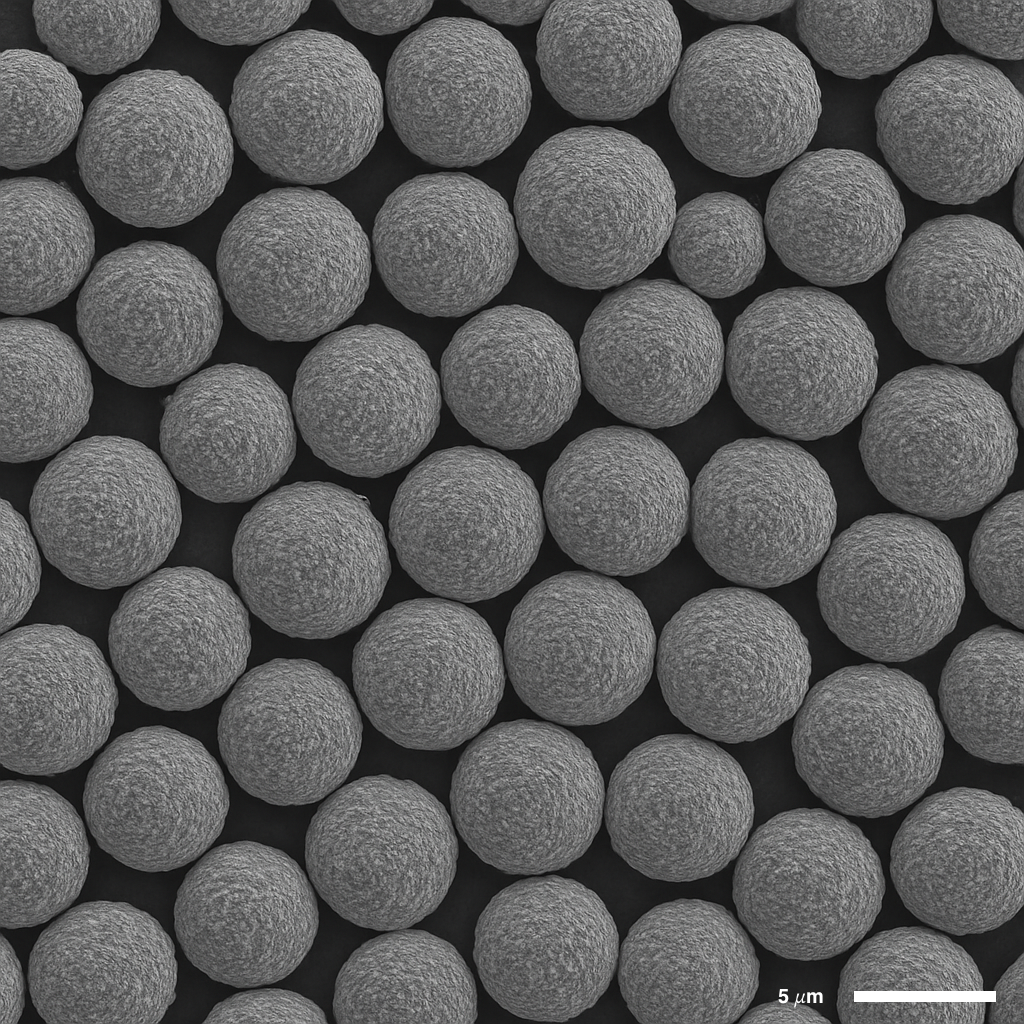}
        \caption{SEM image of NCM precursor particles (Sample A)}
        \label{fig:sample_a}
    \end{subfigure}
    \hfill
    \begin{subfigure}[b]{0.23\textwidth}
        \centering
        \includegraphics[width=\textwidth]{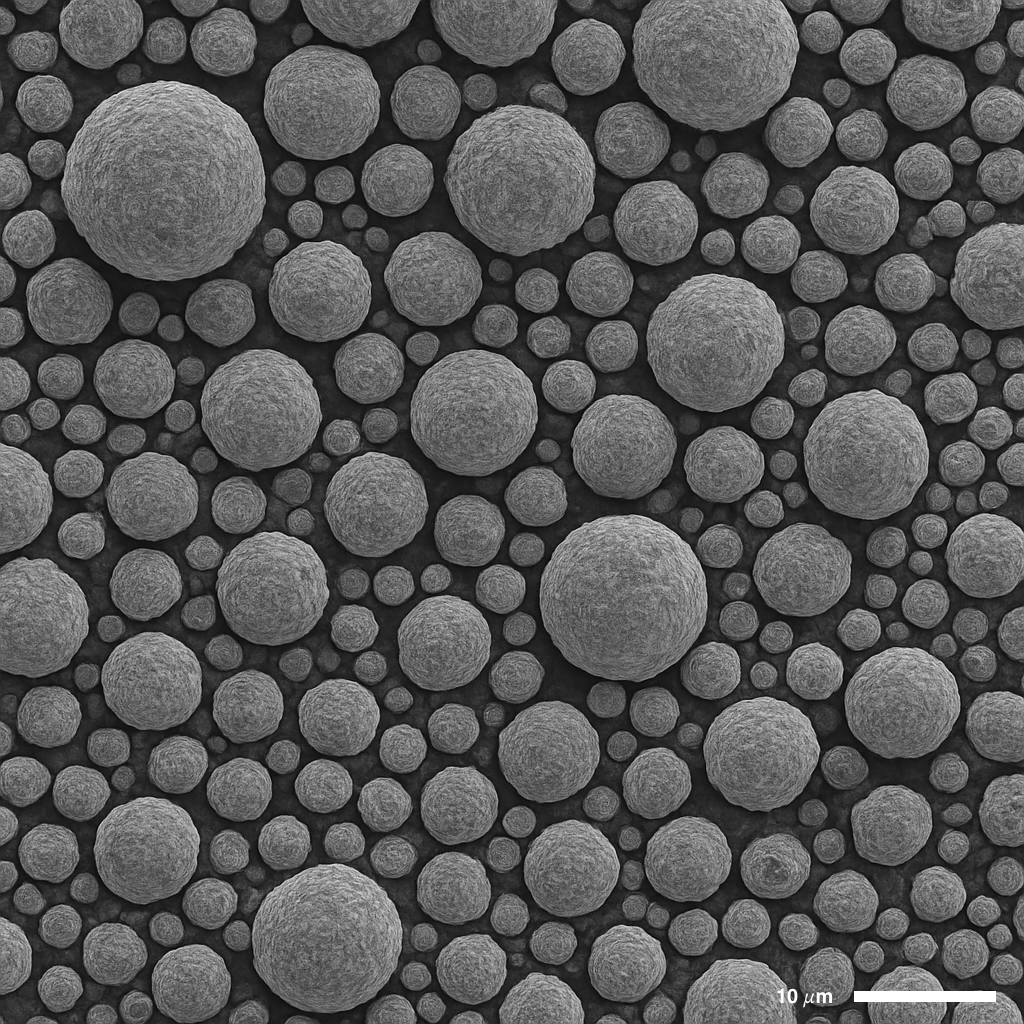}
        \caption{SEM image of NCM precursor particles (Sample B)}
        \label{fig:sample_b}
    \end{subfigure}

    \begin{subfigure}[b]{0.23\textwidth}
        \centering
        \includegraphics[width=\textwidth]{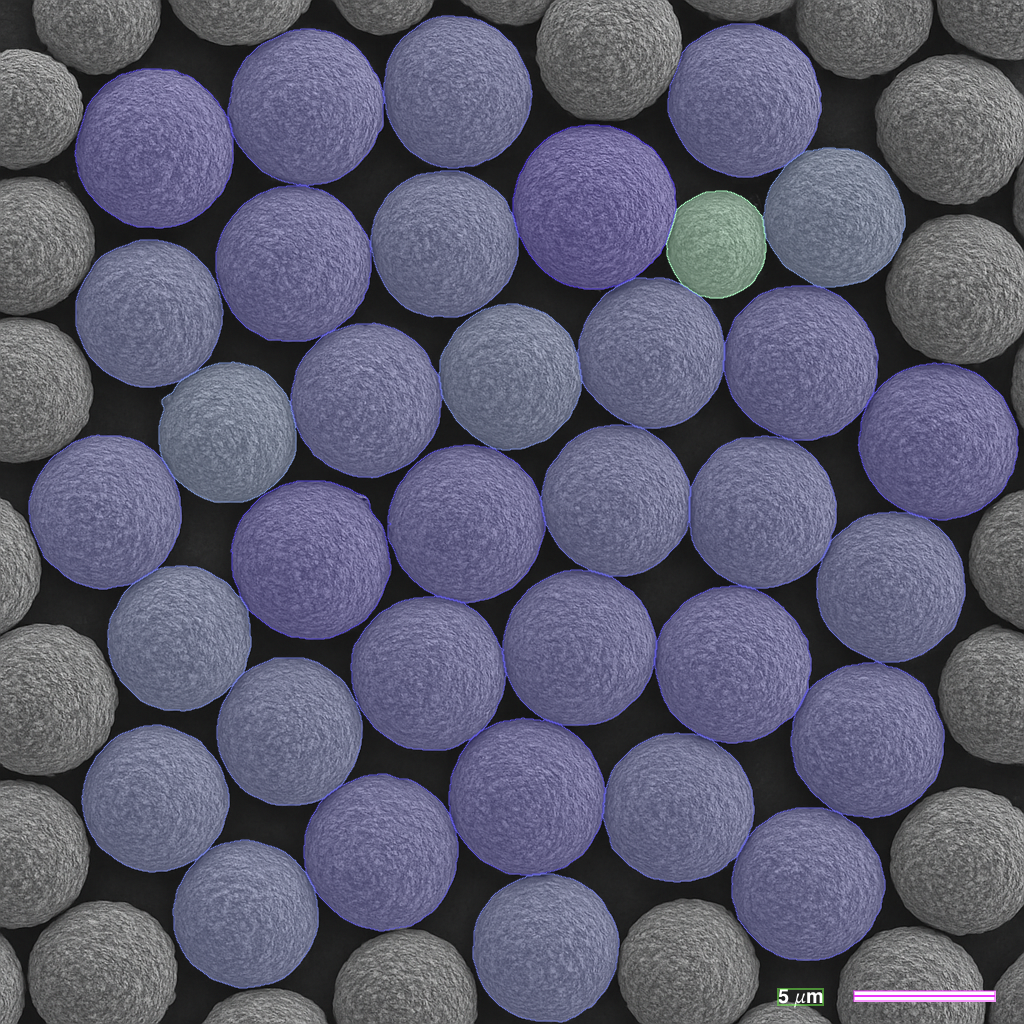}
        \caption{Automatically identified particles and scale bar (Sample A)}
        \label{fig:identified_a}
    \end{subfigure}
    \hfill
    \begin{subfigure}[b]{0.23\textwidth}
        \centering
        \includegraphics[width=\textwidth]{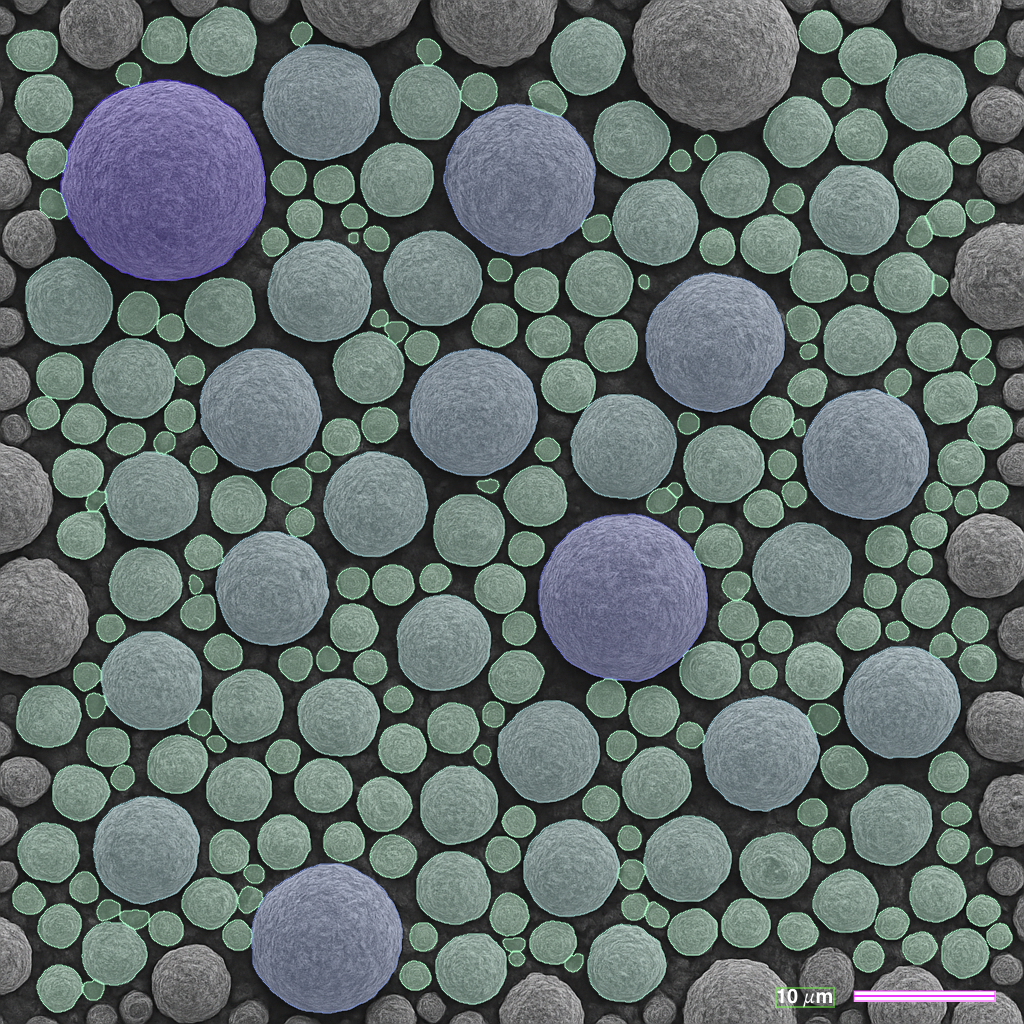}
        \caption{Automatically identified particles and scale bar (Sample B)}
        \label{fig:identified_b}
    \end{subfigure}

    \begin{subfigure}[b]{0.23\textwidth}
        \centering
        \includegraphics[width=\textwidth]{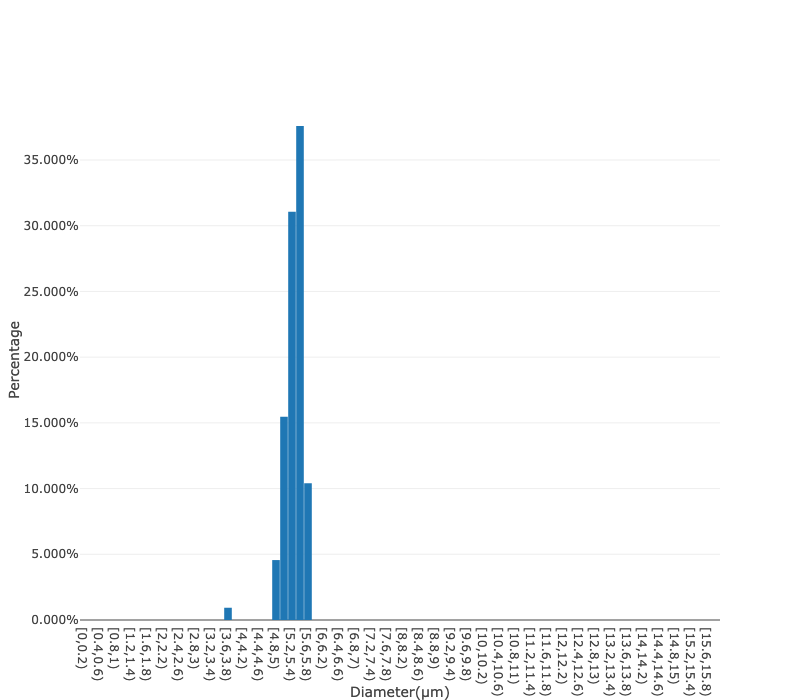}
        \caption{Particle diameter distribution (Sample A)}
        \label{fig:diameter_dist_a}
    \end{subfigure}
    \hfill
    \begin{subfigure}[b]{0.23\textwidth}
        \centering
        \includegraphics[width=\textwidth]{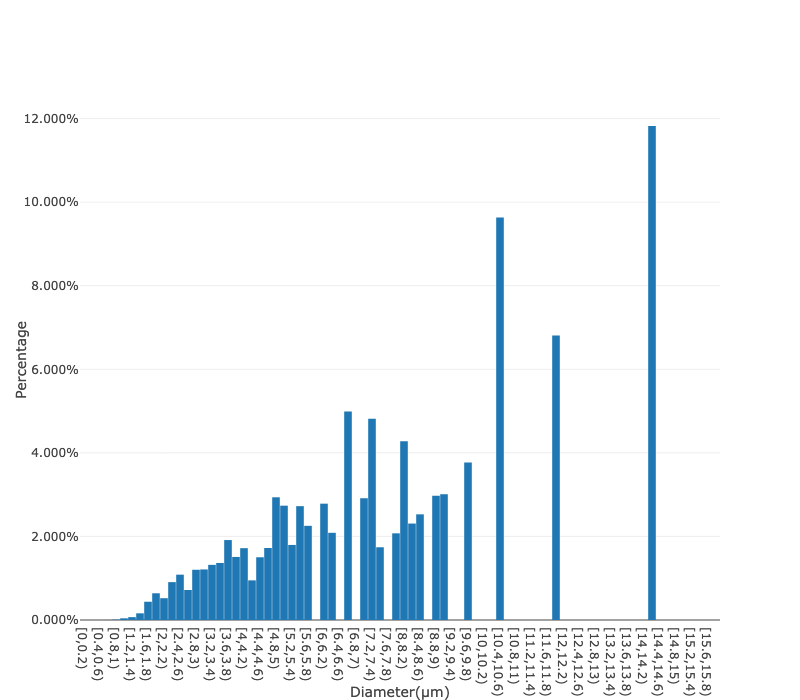}
        \caption{Particle diameter distribution (Sample B)}
        \label{fig:diameter_dist_b}
    \end{subfigure}

    \begin{subfigure}[b]{0.23\textwidth}
        \centering
        \includegraphics[width=\textwidth]{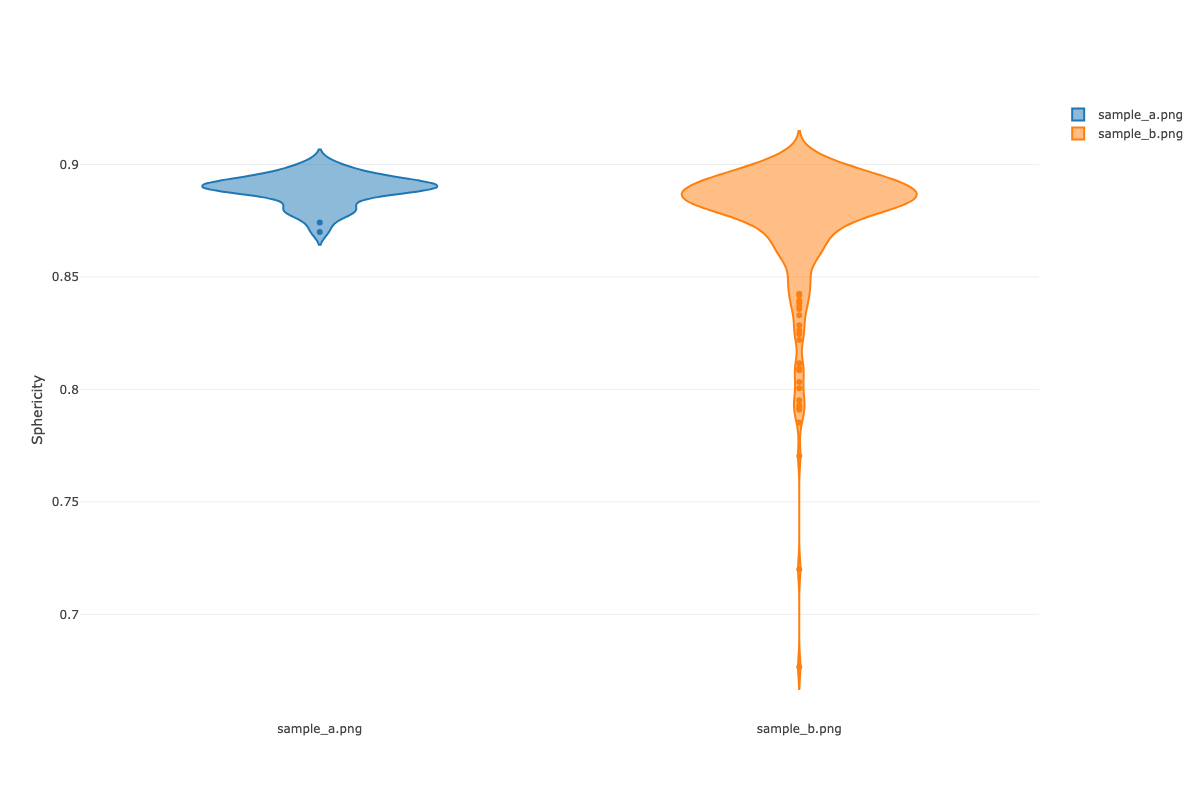}
        \caption{Comparison of sphericity}
        \label{fig:comparison_sphericity}
    \end{subfigure}
    \hfill
    \begin{subfigure}[b]{0.23\textwidth}
        \centering
        \includegraphics[width=\textwidth]{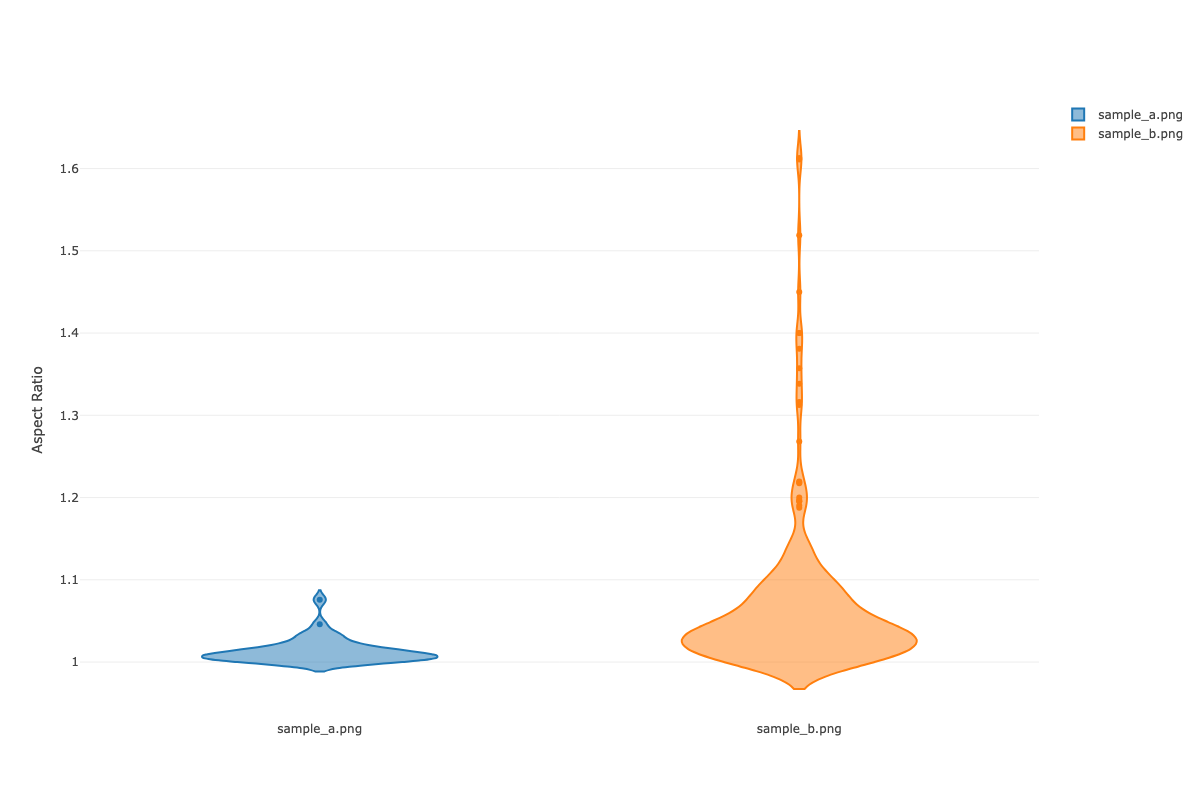}
        \caption{Comparison of aspect ratio}
        \label{fig:comparison_aspect_ratio}
    \end{subfigure}

    \caption{The platform-generated analysis of two NCM precursor samples.}
    \label{fig:application_ncm}
\end{figure}

Figures~\ref{fig:sample_a} and~\ref{fig:sample_b} show the SEM images of NCM precursor particles synthesized under different conditions. Using Figures~\ref{fig:sample_a} and~\ref{fig:sample_b} as inputs for Uni-AIMS, Figures~\ref{fig:identified_a} and~\ref{fig:identified_b} illustrate the particles and scale bars automatically identified by our model. Incomplete particles at the image edges are ignored to ensure accurate particle diameter distribution analysis, which helps to avoid potential biases and inaccuracies that could arise from partially visible particles at the boundaries of the image.The results demonstrate that our platform can quickly and accurately identify precursor particles in electron microscopy images. The size of the two images is \(1024 \times 1024\), and the recognition time on a single 4090 GPU card is approximately 3 seconds. Figures~\ref{fig:diameter_dist_a} and~\ref{fig:diameter_dist_b} present the particle diameter distribution charts provided by the platform. It can be observed that Sample A has a broader particle diameter distribution range, while Sample B has a narrower range. Figures~\ref{fig:comparison_sphericity} and~\ref{fig:comparison_aspect_ratio} show the violin plots comparing the sphericity and aspect ratio of the two samples, also automatically generated by the platform.

\subsection{Assisting Research on the Impact of LiF Microspheres on Battery Performance}

\subsubsection*{Problem Description}

In lithium metal batteries, lithium fluoride (LiF) is considered one of the effective components of the lithium metal interface due to its relatively fast grain boundary conduction capability. However, the mechanism of its action has not yet been fully elucidated. Previous studies have indicated that in fluorinated electrolytes, LiF primarily exists in the form of LiF@Li$_2$O microspheres, which do not directly contact the active material but are instead regarded as a type of free-standing solid electrolyte interphase (SEI) (\cite{2021Revisiting}, \cite{weng2023temperature}). Experimental observations have shown that the decomposition of lithium salts and solvents can both lead to the formation of LiF microspheres. To gain a deeper understanding of the role of LiF in lithium metal batteries and its specific impact on battery performance, it is necessary to conduct precise statistical analysis of the size and quantity of microspheres formed in different electrolytes and correlate these with the deposition/dissolution behavior of metallic lithium.

However, the size of LiF microspheres is typically small, and in some cases, their distribution is relatively dense. This makes manual measurement of the size and quantity of the microspheres extremely challenging. Manual measurement is not only time-consuming and labor-intensive but also prone to statistical errors due to subjective factors. Moreover, manual measurement cannot easily handle large-scale sample analysis, which limits the systematic study of the relationship between the properties of LiF microspheres and battery performance.

\subsubsection*{Solution}

\begin{figure}[htbp]
    \centering
    \begin{subfigure}[b]{0.23\textwidth}
        \centering
        \includegraphics[width=\textwidth]{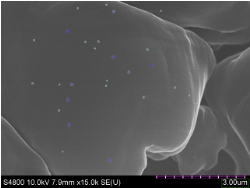}
        \caption{Electrolyte Sample A}
        \label{fig:sub1}
    \end{subfigure}
    \hfill 
    \begin{subfigure}[b]{0.23\textwidth}
        \centering
        \includegraphics[width=\textwidth]{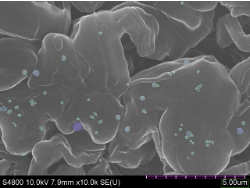}
        \caption{Electrolyte Sample B}
        \label{fig:sub2}
    \end{subfigure}

    \begin{subfigure}[b]{0.23\textwidth}
        \centering
        \includegraphics[width=\textwidth]{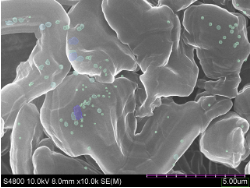}
        \caption{Electrolyte Sample C}
        \label{fig:sub3}
    \end{subfigure}
    \hfill 
    \begin{subfigure}[b]{0.23\textwidth}
        \centering
        \includegraphics[width=\textwidth]{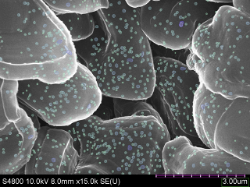}
        \caption{Electrolyte Sample D}
        \label{fig:sub4}
    \end{subfigure}
    \caption{The automatic identification results of LiF microspheres of the SEM images of four different electrolytes by Uni-AIMS.}
    \label{fig:lif}
\end{figure}

To address the aforementioned issues, the Uni-AIMS was applied to the analysis of LiF microspheres. The Uni-AIMS can automatically identify LiF microspheres in different electrolytes in batch mode and efficiently and accurately calculate various metrics according to research requirements, thereby correlating them with performance. Figures~\ref{fig:lif} show the automatic identification results of LiF microspheres of the SEM images of four different electrolytes by Uni-AIMS. It can be seen from the images that Uni-AIMS can comprehensively and accurately identify LiF microspheres in the images. Table~\ref{tab:lif_stats} provides the statistical information of LiF microspheres in the four electrolytes. The application of the Uni-AIMS software significantly improves the efficiency and accuracy of the statistical analysis of LiF microspheres, providing strong support for in-depth research on the role of LiF in lithium metal batteries.

\begin{table}[htbp]
    \centering
    \caption{Statistical information of LiF microspheres in four different electrolytes.}
    \label{tab:lif_stats}
    \begin{tabular}{lcccc}
        \toprule
        Electrolyte & Count & D10($\mu$m)& D50($\mu$m) & D90($\mu$m) \\
        \midrule
        Sample A & 27 & 0.04731 & 0.07047 & 0.1117 \\
        Sample B & 58 & 0.08598 & 0.01325 & 0.2597 \\
        Sample C & 94 & 0.08496 & 0.1424 & 0.2436 \\
        Sample D & 779 & 0.05105 & 0.07314 & 0.09993 \\
        \bottomrule
    \end{tabular}
\end{table}

%% file: Sections/07-Conclusion.tex
\section{Conclusion}

In this work, we propose an intelligent system for the recognition and analysis of microscopy images, featuring high versatility across various types of microscopes and sample categories. Powered by our instance segmentation model, the system achieves precise segmentation of both tiny targets and densely packed particles in microscopy images. With a high degree of automation, the platform enables one-click execution of model inference, deep data mining, visualization, and can automatically generate comprehensive analytical reports. It fully addresses the needs of researchers and industry practitioners for efficient and accurate microstructural analysis. To support model development and system optimization, we have constructed a highly diverse image dataset encompassing a wide range of material types and imaging modalities. Furthermore, we have established a robust multi-source annotation framework, employing a multi-stage process that integrates model-assisted pre-annotation, manual correction, and expert review to ensure the accuracy and consistency of the annotations. 

In future work, we plan to further expand our dataset and enhance the model to improve its generalizability, with the goal of covering a broader range of microscopy scenarios and ultimately enabling the capability to segment and analyze ``anything'' within microscopy images.

On the other hand, to advance the Uni-AIMS platform and align it with emerging demands in automated microscopy analysis, strategic efforts will focus on enhancing user-customizable segmentation capabilities, optimizing real-time performance for complex workflows like FIB (Focused Ion Beam) operations, and deepening integration with automated industrial and research pipelines. These initiatives aim to position Uni-AIMS as a versatile, high-performance tool for next-generation microscopy innovation.

\subsubsection*{Speed Optimization for FIB and High-Throughput Workflows} 
Achieving near-real-time recognition and segmentation (under 10ms  per 1K image) is critical for supporting FIB’s iterative imaging and milling processes. Technical enhancements will focus on model efficiency, including replacing heavy backbones like ViT with lightweight alternatives, optimized for GPU/TPU inference. Post-training quantization (FP16/INT8) and layer pruning will reduce computational load without sacrificing accuracy. Hardware acceleration will involve deploying CUDA-optimized kernels to parallelize pixel dynamics post-processing and implementing multi-scale inference for large-area scans via tiling and seamless stitching. Workflow-specific optimizations, such as preprocessing common FIB imaging patterns for preemptive caching and dynamically adjusting resolutions during navigation phases, will further reduce latency. Validation will measure end-to-end workflow latency (image acquisition to feedback loops). These optimizations enable real-time feedback during precision milling without throughput compromises.

\subsubsection*{Integration with Automated Workflows and Standards} 
Uni-AIMS will deepen compatibility with industrial and research pipelines by expanding its API ecosystem, standardizing reporting, and fostering collaboration. RESTful APIs will be extended to support batch processing, equipment integration (e.g., SEM, FIB-SEM), and Python SDKs for custom scripting. Automated reporting tools will generate ISO/GBT-compliant analyses (e.g., GBT 6394-2017 for grain boundary sizing) with uncertainty quantification, while third-party certifications will validate the platform for regulated sectors like aerospace and energy. Collaborative features, including version-controlled project sharing and multi-user annotation tools, will streamline team-based research. These efforts strengthen adoption in industrial QA/QC pipelines, particularly for semiconductor defect characterization and battery R\&D, positioning Uni-AIMS as a gold-standard tool for compliance-driven applications.